\documentclass[aps,prd,longbibliography,preprintnumbers,nofootinbib,
superscriptaddress,amsmath,floatfix,10pt]{revtex4}

\usepackage{amssymb}  
\usepackage{amsmath}
\usepackage{hyperref}
\usepackage{graphicx
}
\usepackage{exscale}
\usepackage{bbold}
\usepackage{textcomp}
\usepackage{enumerate}
\usepackage [english]{babel}
\usepackage[utf8]{inputenc}
\usepackage [autostyle, english = american]{csquotes}
\MakeOuterQuote{"}

\usepackage{color}

\usepackage[normalem]{ulem}  

\newcommand{\non}{\nonumber\\}

\newcommand{\be}{\begin{equation}}
\newcommand{\ee}{\end{equation}}
\newcommand{\bea}{\begin{eqnarray}}
\newcommand{\eea}{\end{eqnarray}}
\newcommand{\ba}[1]{\begin{array}{#1}}
\newcommand{\ea}{\end{array}}

\begin{document}

\title{
How neutron star properties disfavor a nuclear chiral density wave
}

\author{Orestis Papadopoulos}
\email{op1g19@soton.ac.uk}
\affiliation{Mathematical Sciences and STAG Research Centre, University of Southampton, Southampton SO17 1BJ, United Kingdom}

\author{Andreas Schmitt}
\email{a.schmitt@soton.ac.uk}
\affiliation{Mathematical Sciences and STAG Research Centre, University of Southampton, Southampton SO17 1BJ, United Kingdom}

\date{21 January 2025}


\begin{abstract} 
Cold and dense matter may break rotational symmetry spontaneously and thus form an anisotropic phase in the interior of neutron stars. We consider the concrete example of an anisotropic chiral condensate in the form of a chiral density wave. Employing a nucleon-meson model and taking into account fermionic vacuum fluctuations, we improve and extend previous results by imposing the conditions of electric charge neutrality and electroweak equilibrium, by allowing for a more general form of the vector meson self-interactions, and by including properties of pure neutron matter into the fit of the model parameters. We find that the conditions inside neutron stars postpone the onset of the chiral density wave to larger densities compared to isospin-symmetric nuclear matter. While this still allows for the construction of stars with an anisotropic core, we find that the chiral density wave is energetically preferred only in a corner of the parameter space where matter is too soft to generate  stars with realistic masses.  Therefore, taking into account constraints from astrophysical data, our calculation predicts an isotropic neutron star core.

\end{abstract}

\maketitle


\section{Introduction}
\label{sec:intro}

The phase structure  of Quantum Chromodynamics (QCD) at low temperatures and large, but not asymptotically large, baryon densities is poorly understood. Numerous candidate phases have been suggested, including phases that spontaneously break rotational and/or translational invariance. Anisotropies or inhomogeneities are generated for instance through spatial modulations of the chiral condensate \cite{Dautry:1979bk,Takahashi:2001jq,Takahashi:2002md,Buballa:2014tba} or the Cooper pair condensate \cite{Alford:2000ze,Schafer:2005ym,Kryjevski:2005qq,Alford:2007xm}.  
Locating an anisotropic or inhomogeneous phase in the QCD phase diagram of temperature and baryon chemical potential does not necessarily translate into the existence of this phase in the interior of a neutron star, currently our only ``laboratory'' for ultra-dense matter. The reason is that, firstly, neutron star conditions require a nonzero isospin chemical potential and, secondly, it has to be calculated if the density within a realistic neutron star is sufficiently large to accommodate the predicted anisotropic phase. In this paper we are concerned with this question on the level of a phenomenological model: We implement neutron star conditions into a nucleon-meson model, where an anisotropic chiral condensate in the form of a chiral density wave (CDW) was predicted for isospin-symmetric nuclear matter \cite{Pitsinigkos:2023xee}, and investigate whether the CDW can be found in a realistic star. 

In a CDW (also known as ``dual chiral density wave'', ``chiral spiral'', or ``axial wave condensation'') the chiral condensate oscillates between scalar and pseudoscalar sectors, characterized by a single wave vector, whose 
magnitude is dynamically determined. As a consequence, the fermionic dispersion relations become anisotropic. 
The CDW or one of its variants is expected to occur in the vicinity of the chiral transition \cite{Nickel:2009wj}, even if this transition is a smooth crossover \cite{Pitsinigkos:2023xee}. Therefore, it is important to work with a model that incorporates chiral symmetry and the spontaneous breaking thereof. Here, we employ a model based on nucleonic degrees of freedom, which interact via meson exchange \cite{Boguta:1982wr,Boguta:1986ha,Floerchinger:2012xd,Drews:2013hha,Drews:2014spa,Fraga:2018cvr,Schmitt:2020tac,Fraga:2022yls,Pitsinigkos:2023xee}. The parameters of the model are chosen to reproduce properties of low-density nuclear matter. Importantly, the nucleon mass is generated purely dynamically. Therefore, this model has a well-defined chiral limit (setting the pion mass to zero), which we include as a benchmark for our main results with a physical pion mass.
In the chiral limit, massless fermions exist in the high-density phase. These acquire a small mass once a realistic explicit chiral symmetry breaking is included. The high-density phase 
can be viewed as a toy version of quark matter or -- perhaps somewhat exotic but not excluded by first-principle arguments -- as a confined but chirally symmetric phase of QCD. In this sense, our approach is complementary to phenomenological models based on quark degrees of freedom, such as the quark-meson or Nambu--Jona-Lasinio (NJL) models, where it is the low-density phase which is a toy version of real-world QCD due to the lack of realistic nuclear matter. It is in these models that the CDW  has been studied most frequently  \cite{Nakano:2004cd,Nickel:2009wj,Frolov:2010wn,Carignano:2011gr,Carignano:2014jla,Buballa:2015awa,Adhikari:2017ydi,Andersen:2018osr,Ferrer:2019zfp,Carignano:2019ivp,Lakaschus:2020caq,Buballa:2020xaa},  while there are much fewer studies of the  nuclear CDW \cite{Dautry:1979bk,Takahashi:2001jq,Takahashi:2002md,Heinz:2013hza,Takeda:2018ldi,Pitsinigkos:2023xee}. In this paper we will not be concerned with more complicated spatial structures, which can be different one-dimensional modulations, or three-dimensional crystalline structures \cite{Abuki:2011pf,Carignano:2012sx,Buballa:2014tba,Takeda:2018ldi,Pisarski:2020dnx,Winstel:2024qle}. Another class of inhomogeneous phases, also not considered here, arises from spatial separation of two different phases, for instance a quark-hadron mixed phase \cite{Glendenning:1992vb,PhysRevLett.70.1355,Schmitt:2020tac}.

Compact stars containing a CDW phase were previously constructed in an NJL model; in a pure quark star (and ignoring  electric charge neutrality) \cite{Buballa:2015awa} and combined with isotropic nuclear matter from a different model, in the presence of a magnetic field  \cite{Carignano:2015kda}. External fields -- not included in our present study -- that explicitly break 
rotational symmetry  tend to favor anisotropic phases such as the CDW \cite{Frolov:2010wn,Ferrer:2021mpq}. The (quark) CDW in the context of neutron stars was also considered in Ref.\ \cite{Tatsumi:2014cea}, where the neutrino emissivity was computed, however without checking whether a stable star with a CDW is realized. Our main novelty compared to these works is that we construct stars with a potential CDW core from a single model that respects low-density nuclear matter properties and basic neutron star conditions, while at the same time allowing for a chiral transition and a dynamical calculation of whether the CDW is energetically preferred or not.

Our work builds on the recent study of Ref.\ \cite{Pitsinigkos:2023xee}. The model used in this reference is extended by including rho meson exchange and a more general form of the vector meson self-interactions. This allows us to include properties of pure neutron matter, known from chiral effective field theory ($\chi$EFT) \cite{Tews:2018kmu}, into the parameter fit, following the idea of Ref.\ \cite{Alford:2022bpp}, where a similar model was considered. Also, we add electrons and muons and implement the conditions of electric charge neutrality and equilibrium with respect to the electroweak interactions. We evaluate the model at zero temperature, neglecting mesonic fluctuations, but, importantly, include the nucleonic Dirac sea, which receives nontrivial contributions from the CDW. We will employ the renormalization scheme introduced in Ref.\ \cite{Pitsinigkos:2023xee}, which ensures boundedness of the effective potential in the direction of the CDW wavenumber and thus alleviates artifacts encountered previously in similar models with quark degrees of freedom \cite{Carignano:2011gr,Carignano:2014jla}.

Our paper is organized as follows. We explain the model and our approximations in Sec.\ \ref{sec:setup}. This includes the CDW ansatz,  Sec.\ \ref{sec:Lag}, the derivation of the free energy density and the equation of state under neutron star conditions, Sec.\ \ref{sec:free}, the stationarity equations for the meson condensates, Sec.\ \ref{sec:stat}, and the procedure for fitting our parameters, Sec.\ \ref{sec:para}. Our results are presented in Sec.\ \ref{sec:results}, where we start with the results for the phase diagram, Sec.\ \ref{sec:phase}, and the mass-radius curves for neutron stars, Sec.\ \ref{sec:MR}, before we present and discuss our main results regarding the CDW in neutron stars in Secs.\ \ref{sec:locate} and \ref{sec:discussion}. We summarize our results and give an outlook in Sec.\ \ref{sec:summary}. 

\section{Setup}
\label{sec:setup}

\subsection{Lagrangian and CDW ansatz}
\label{sec:Lag}

Our model describes neutrons and protons with dynamically generated mass, interacting via scalar and vector meson exchange. This model was employed in different variants previously; for instance, for isospin-symmetric nuclear matter \cite{Fraga:2018cvr,Pitsinigkos:2023xee}, with a simpler \cite{Pitsinigkos:2023xee} or without \cite{Schmitt:2020tac} meson self-interaction, and including strangeness \cite{Fraga:2022yls}. The Lagrangian has a nucleonic, a mesonic, and an interaction part,  
\be
{\cal L} = {\cal L}_{\rm nucl} + {\cal L}_{\rm mes} + {\cal L}_{\rm int} \, . 
\ee
The nucleonic part is given by
\bea
{\cal L}_{\rm nucl}  = \bar{\psi}(i\gamma^\mu\partial_\mu + \gamma^0\hat{\mu})\psi \, , 
\eea
where 
\be
\psi = \left(\begin{array}{c}\psi_n \\ \psi_p\end{array}\right) 
\ee
includes neutron and proton spinors, where $\bar{\psi} = \psi^\dag \gamma^0$, and where  
\be \label{muhat}
\hat{\mu}=\left(\begin{array}{cc} \mu_n & 0 \\ 0 & \mu_p \end{array}\right)  
= \left(\begin{array}{cc} \mu_B+\mu_I & 0 \\ 0 & \mu_B-\mu_I \end{array}\right)
\ee
is the chemical potential matrix in isospin space with neutron  and proton chemical potentials $\mu_n$, $\mu_p$, which can be expressed in terms of baryon and isospin chemical potentials $\mu_B$ and $\mu_I$. 
The mesonic part ${\cal L}_{\rm mes}$ includes the scalar and pseudoscalar meson fields $\sigma$ and $\pi = \pi_a\tau^a$, where $\tau^a$ are the Pauli matrices, and the vector meson fields $\omega^\mu$ and $\rho^\mu = \rho^\mu_a\tau^a$, 
\bea
{\cal L}_{\rm mes}  &=& \frac{1}{2}\partial_\mu\sigma\partial^\mu\sigma + \frac{1}{2}\partial_\mu\pi_a\partial^\mu\pi^a - {\cal U}(\sigma,\vec{\pi})
 \non[2ex]
&& +\frac{1}{2}m_\omega^2\omega_\mu\omega^\mu+\frac{1}{2}m_\rho^2\rho_\mu^3\rho^\mu_3 +\frac{d_\omega}{4}(\omega_\mu\omega^\mu)^2 +\frac{d_\rho}{4}(\rho_\mu^3\rho^\mu_3)^2+\frac{d_{\omega\rho}}{2}\omega_\mu\omega^\mu\rho_\nu^3\rho^\nu_3+ \ldots \, . 
\eea
We have, for notational simplicity, written down an effective mesonic Lagrangian, where we have omitted derivatives of the vector meson fields and the $a=1,2$ components of the rho meson, indicated by the dots. This is sufficient for our purposes because we will only allow for nonzero background fields in the temporal components,
\be
\omega\equiv \langle \omega_0\rangle \, , \qquad 
\rho\equiv \langle \rho_0^3\rangle \, , 
\ee
which we assume to be constant in space and time, and neglect all mesonic fluctuations\footnote{Spatial vector meson condensates $\langle \omega_i\rangle$, $\langle \rho_i^3\rangle$ vanish due to the absence of spatial baryon currents \cite{glendenning}. This is not changed by the CDW.}. 
We will work with the vector meson masses $m_\omega\simeq m_\rho \simeq 782\, {\rm MeV}$. We have allowed for three different quartic self-coupling terms of the vector mesons with coupling constants $d_\omega$, $d_\rho$, $d_{\omega\rho}$; see also Ref.\ \cite{Steiner:2004fi}, where, in a similar model, additional, higher-order mesonic interactions terms were included. The derivative terms of the $\sigma$ and $\pi$ fields cannot be dropped, even though we will neglect their fluctuations as well. The reason is that these terms will contribute due to the CDW. The potential of the scalar mesons is 
\be
{\cal U}(\sigma,\vec{\pi}) = \sum_{n=1}^4 \frac{a_n}{n!} \frac{(\sigma^2+\pi_a\pi^a-f_\pi^2)^n}{2^n}-\epsilon(\sigma-f_\pi) \, , 
\ee
with parameters $\epsilon, a_1, a_2, a_3, a_4$ and the pion decay constant $f_\pi\simeq 92\, {\rm MeV}$. In the presence of a nonzero parameter $\epsilon$, chiral symmetry is broken explicitly. Finally, the nucleon-meson interaction is given by 
\bea
{\cal L}_{\rm int} &=& 
 - \bar{\psi}\Big[g_\sigma(\sigma+ i\gamma^5\pi) + \gamma^\mu(g_\omega\omega_\mu+g_\rho\rho_\mu) \Big]\psi \, , 
\eea
 with the Yukawa coupling constants $g_\sigma$, $g_\omega$, and $g_\rho$. 

The CDW is included by the ansatz
\be \label{CDW}
\langle \sigma \rangle  = \phi\cos(2\vec{q}\cdot\vec{x}) \, , \qquad \langle \pi_3\rangle  = \phi\sin(2\vec{q}\cdot\vec{x}) \, , \qquad \langle \pi_1\rangle = \langle \pi_2\rangle = 0 \, , 
\ee
where $\phi$ and $\vec{q}$ do not depend on space and time and will later be determined dynamically. The direction of $\vec{q}$ is fixed but arbitrary and breaks rotational symmetry. The form (\ref{CDW}) of the sigma and neutral pion condensates solves the equations of motion for the scalar fields in the absence of nucleons and in the chiral limit $\epsilon=0$. In the realistic case of nonzero $\epsilon$ the CDW is no longer a solution even in the absence of nucleons. Therefore, Eq.\ (\ref{CDW}) should be considered as an ansatz which may or may not reduce the free energy of the nucleon-meson system. Variations and refinements of this ansatz include a spatial modulation only in the $\sigma$ direction in terms of Jacobi elliptic functions \cite{Nickel:2009wj}, a shift of the circular modulation in the $\sigma$-$\pi_3$ plane  \cite{Takeda:2018ldi}, or allowing for charged pion condensates \cite{Dautry:1979bk}. 
The CDW ansatz allows us to simplify the calculation with the help of the field transformation  
\be
\psi \to e^{-i\gamma^5\tau_3 \vec{q}\cdot\vec{x}} \psi \, .
\ee
Then, neglecting all mesonic fluctuations, we obtain the effective ``mean-field'' Lagrangian 
\bea \label{Leff}
{\cal L}_{\rm mf} &=& 
\bar{\psi}(i\gamma^\mu\partial_\mu + \gamma^0\hat{\mu}^*-M+\gamma^5\vec{q}\cdot\vec{\gamma}\tau_3)\psi -V(\omega,\rho)  - U(\phi) - \Delta U(\phi,q)  \, , 
\eea
with the effective nucleon mass
\be
M = g_\sigma \phi \, ,
\ee
and $\hat{\mu}^* = {\rm diag}(\mu_n^*,\mu_p^*)$ with the effective neutron and proton chemical potentials, 
\begin{subequations} \label{mustar}
\bea
\mu_n^* &=& \mu_n-g_\omega\omega-g_\rho\rho \, , \\[2ex]
\mu_p^*&=&  \mu_p-g_\omega\omega+g_\rho\rho \, .
\eea
\end{subequations}
The vector meson potential is 
\bea
V(\omega,\rho) = - \frac{m_\omega^2}{2}\omega^2- \frac{m_\rho^2}{2}\rho^2 - \frac{d_\omega}{4}\omega^4 - \frac{d_\rho}{4}\rho^4 - \frac{d_{\omega\rho}}{2}\omega^2\rho^2 \, , 
\eea
while the scalar meson potential is given by 
\bea \label{Uphiq}
U(\phi) = \sum_{n=1}^4 \frac{a_n}{n!} \frac{(\phi^2-f_\pi^2)^n}{2^n}-\epsilon(\phi-f_\pi)  \, , \qquad \Delta U(\phi,q) =  2\phi^2q^2 +(1-\delta_{q0})\epsilon\phi  \, .
\eea
The contribution $\Delta U$  contains the kinetic energy $2\phi^2q^2$ due to the   axial current in the mesonic sector, and we have employed a spatial average over the explicit symmetry breaking term, which induces a discontinuity \cite{Pitsinigkos:2023xee}: While in the chiral limit the CDW solution can connect continuously to the isotropic solution, this is not the case for nonzero $\epsilon$. As a consequence, the phase transition between isotropic and anisotropic phases cannot be continuous in the physical case.

\subsection{Free energy density}
\label{sec:free}

The effective Lagrangian (\ref{Leff}) is quadratic in the fermionic fields and otherwise only contains classical background fields. Therefore, we can compute the free energy density without further approximations. To this end, we first identify  the inverse nucleon propagator in momentum space,
\be \label{SK}
S^{-1}(K) = -\gamma^\mu K_\mu + M -\hat{\mu}^*\gamma^0+\vec{q}\cdot\vec{\gamma}\gamma^5\tau_3 \, , 
\ee
where $K^\mu = (k_0,\vec{k})$ with $k_0 = -i\omega_n$ and the fermionic Matsubara frequencies $\omega_n = (2n+1)\pi T$, where $n\in \mathbb{Z}$ and $T$ is the temperature. For completeness we derive the free energy for arbitrary temperatures, although for the purpose of this paper it is sufficient to evaluate the model at $T=0$. In any case, our mean-field approximation of neglecting the mesonic fluctuations cannot be expected to be very accurate at large temperatures. 

The inverse propagator (\ref{SK}) is diagonal in flavor space such that we can write
\be
S^{-1} = {\rm diag}[S_0^{-1}(\mu_n^*,\vec{q}\,),S_0^{-1}(\mu_p^*,-\vec{q}\,)]\, , 
\ee
where 
\be
S_0^{-1}(\mu,\vec{q}\,) \equiv -\gamma^\mu K_\mu + M -\mu\gamma^0+\vec{q}\cdot\vec{\gamma}\gamma^5 \, . 
\ee
Inversion gives the propagator 
\be
S= {\rm diag}[S_0(\mu_n^*,\vec{q}\,),S_0(\mu_p^*,-\vec{q}\,)] \, , 
\ee
with 
\be
S_0(\mu,\vec{q}\,)  \equiv \frac{(-\gamma^\mu K_\mu-M-\mu\gamma^0+\vec{q}\cdot\vec{\gamma}\,\gamma^5)(-\gamma^\mu K_\mu+M-\mu\gamma^0-\vec{q}\cdot\vec{\gamma}\,\gamma^5)
(-\gamma^\mu K_\mu-M-\mu\gamma^0-\vec{q}\cdot\vec{\gamma}\,\gamma^5)}{[(k_0+\mu)^2-(E_k^+)^2][(k_0+\mu)^2-(E_k^-)^2]} \, ,
\ee
where 
\be
E_k^\pm = \sqrt{\left(\sqrt{k_\ell^2+M^2}\pm q\right)^2+k_\perp^2} 
\ee 
are the single-nucleon dispersion relations with longitudinal and transverse components of the momentum $\vec{k}_{\ell} = \hat{\vec{q}}\,\hat{\vec{q}}\cdot\vec{k}$, $\vec{k}_\perp = \vec{k}-\hat{\vec{q}}\,\hat{\vec{q}}\cdot\vec{k}$. The dispersion relations are the poles of the propagator or, in other words, the zeros of the determinant of the inverse propagator,
\be
{\rm det}\, S^{-1} = [(k_0+\mu_n^*)^2-(E_k^+)^2][(k_0+\mu_n^*)^2-(E_k^-)^2][(k_0+\mu_p^*)^2-(E_k^+)^2][(k_0+\mu_p^*)^2-(E_k^-)^2] \, . 
\ee
The free energy is obtained from the logarithm of the partition function, following the standard calculation in imaginary-time thermal field theory. After performing the sum over Matsubara frequencies the nucleonic contribution to the free energy is
\bea
\Omega_{\rm nucl} &=& - \sum_{N=n,p}\sum_{e=\pm}\sum_{s=\pm} \int\frac{d^3\vec{k}}{(2\pi)^3}\left\{\frac{E_k^s}{2}+T\ln\left[1+e^{-(E_k^s-e\mu_N^*)/T}\right]\right\} = -P_{\rm vac} - P_{\rm mat}\, ,
\eea
where we have denoted the vacuum  contribution (``Dirac sea'') to the pressure by $P_{\rm vac}$ and the matter contribution of both nucleonic degrees of freedom by $P_{\rm mat}$.
In the zero-temperature limit, anti-nucleons do not contribute to the matter part, which becomes 
\bea \label{Pmatdef}
P_{\rm mat}  &=& \frac{1}{2\pi^2}\sum_{N=n,p}\sum_{s=\pm} \int_0^\infty dk_\ell\int_0^\infty dk_\perp k_\perp (\mu_N^*-E_k^s)\Theta(\mu_N^*-E_k^s) \non[2ex]
&=&\sum_{N=n,p} {\cal P}(M,q,\mu_N^*) \, ,
\eea
with
\bea \label{calD} \allowdisplaybreaks
{\cal P}(M,q,\mu)&\equiv &\frac{\Theta(\mu-q-M)}{16\pi^2}\left\{M^2[M^2+4q(q-\mu)]\ln\frac{\mu-q+k_-}{M}+\frac{k_-}{3}[2(\mu^2-q^2)(\mu-q)-M^2(5\mu-13q)]\right\} \non[2ex]
&&+\frac{\Theta(\mu+q-M)}{16\pi^2}\left\{M^2[M^2+4q(q+\mu)]\ln\frac{\mu+q+k_+}{M}+\frac{k_+}{3}[2(\mu^2-q^2)(\mu+q)-M^2(5\mu+13q)]\right\} \non[2ex]
&&+\frac{\Theta(q-\mu-M)}{16\pi^2}\left\{M^2[M^2+4q(q-\mu)]\ln\frac{q-\mu+k_-}{M}-\frac{k_-}{3}[2(\mu^2-q^2)(\mu-q)-M^2(5\mu-13q)]\right\} \non[2ex]
&&-\frac{\Theta(q-M)}{8\pi^2}\left[M^2(M^2+4q^2)\ln\frac{q+\sqrt{q^2-M^2}}{M}-\frac{\sqrt{q^2-M^2}}{3}q(2q^2+13M^2)\right] \, ,
\eea
where we have abbreviated 
\begin{equation} \label{kpm}
k_\pm\equiv \sqrt{(\mu\pm q)^2-M^2} \, .
\end{equation}
The vacuum part $P_{\rm vac}$ is exactly the same as for isospin symmetric matter since the effective masses of neutron and proton are degenerate and the chemical potentials do not enter the vacuum contribution. Therefore, we can exactly follow the renormalization procedure of Ref.\ \cite{Pitsinigkos:2023xee}. As a result, the contribution of the Dirac sea can be combined with the scalar potential to obtain a modified potential with renormalized parameters (for which we use the same symbols as for the bare parameters), 
\begin{subequations}
\bea \label{Utilde}
\tilde{U}(\phi) &\equiv&  U(\phi)+\frac{m_N^4}{96\pi^2}\left(1-8\frac{\phi^2}{f_\pi^2}-12\frac{\phi^4}{f_\pi^4}\ln\frac{\phi^2}{f_\pi^2} +8\frac{\phi^6}{f_\pi^6} -\frac{\phi^8}{f_\pi^8}\right) \, , \\[2ex]
\label{DelUtilde}
\Delta\tilde{U}(\phi,q) &\equiv&  \Delta U(\phi,q)-\frac{q^2M^2}{2\pi^2}\ln\frac{M^2}{\ell^2} -\frac{q^4}{2\pi^2}F(y)\, ,
\eea
\end{subequations}
where $m_N = 939\, {\rm MeV}$ is the vacuum mass of the nucleons, where 
\begin{equation}\label{Fdef}
F(y) \equiv \frac{1}{3}+\Theta(1-y)\left[-\sqrt{1-y^2}\frac{2+13y^2}{6}+2y^2\left(1+\frac{y^2}{4}\right)\ln\frac{1+\sqrt{1-y^2}}{y} \right] \, ,
 \end{equation}
 with 
 \be \label{ydef}
y\equiv \frac{M}{q} \, , 
\ee
and where the renormalization scale is
\be \label{ell}
\ell = \sqrt{m_N^2 + (2cq)^2} \, .
\ee
It was argued in Ref.\ \cite{Pitsinigkos:2023xee} that the $q$ dependence in the renormalization scale is necessary to avoid an unbounded effective potential in the $q$ direction. Implicitly, the renormalization scale now depends on the medium, because $q$ depends on the chemical potential. This is similar to perturbative QCD calculations, where the scale typically is assumed to be a function of temperature and/or chemical potential
\cite{kapustabook,Kurkela:2009gj,Fraga:2023cef}. As a consequence of this choice, curious re-entrance phenomena of the CDW, previously discovered in similar models \cite{Carignano:2011gr,Carignano:2014jla}, can be avoided. 
Without further $q$-dependent renormalization conditions, the renormalization scale is not uniquely fixed and we 
parameterize this freedom by the parameter $c$. All results of Ref.\ \cite{Pitsinigkos:2023xee} were, for simplicity,  obtained with $c=1$, while it was already pointed out in that reference that the minimal $c$, below which the effective potential becomes unbounded, is 
\be \label{c0}
c_0 = e^{-1-2\pi^2/g_\sigma^2} \, .
\ee
In our numerical evaluation, we will explore the dependence of the results on $c$ and in particular cover the case $c=c_0$, which makes the CDW more favorable towards lower chemical potentials. 

Finally, since we are interested in neutron star conditions, we need to add a leptonic contribution to the free energy from electrons and muons, 
\be
\Omega_{\rm lep} = -p(\mu_e,m_e)-p(\mu_\mu,m_\mu) \, ,
\ee
with the electron mass $m_e\simeq 511\, {\rm keV}$, the muon mass $m_\mu\simeq 106\, {\rm MeV}$, electron and muon chemical potentials $\mu_e$ and $\mu_\mu$, and the zero-temperature pressure of a non-interacting Fermi gas
\be
p(\mu,m) = \frac{\Theta(\mu-m)}{8\pi^2}\left[\left(\frac{2}{3}k_F^3-m^2k_F\right)\mu+m^4\ln\frac{k_F+\mu}{m}\right] \, , 
\ee
where $k_F=\sqrt{\mu^2-m^2}$ is the Fermi momentum. 

Putting everything together, we arrive at the zero-temperature free energy 
\be \label{OmPmat}
\Omega = -P_{\rm mat} +V  + \tilde{U} + \Delta \tilde{U} +\Omega_{\rm lep}  \, .
\ee
Up to this point, the derivation is a straightforward generalization of Ref.\ \cite{Pitsinigkos:2023xee} to 
isospin-asymmetric matter -- through different chemical potentials for neutrons and protons and the interaction with rho mesons --  plus the trivial addition of leptons. As a result, we have obtained a free energy that depends on the chemical potentials $\mu_n,\mu_p,\mu_e,\mu_\mu$. In an isolated neutron star, matter is in equilibrium with respect to electroweak reactions \cite{Schmitt:2010pn}. We thus require equilibrium with respect to electron capture  and beta decay,
\begin{subequations} \label{npe}
\bea
p+e&\to& n + \nu_e \\[2ex]
n&\to& p+e+\bar{\nu}_e \, ,
\eea
\end{subequations}
 which, at sufficiently small temperatures, where the neutrino mean free path is of the order of or larger than the size of the system, implies $\mu_p = \mu_n-\mu_e$. The processes (\ref{npe}) also occur with electron and electron neutrino replaced by muon and muon neutrino, which gives the additional constraint $\mu_e=\mu_\mu$. The system is then also, without any further constraint, in equilibrium with respect to the purely leptonic processes
 \begin{subequations}
\bea
e&\to& \mu+\bar{\nu}_\mu+\nu_e \, , \\[2ex]
\mu&\to& e+\bar{\nu}_e+\nu_\mu \, . 
\eea
\end{subequations}
The two constraints allow us to reduce the number of independent chemical potentials to 2, say $\mu_n$ and $\mu_e$.

\subsection{Stationarity equations}
\label{sec:stat}

For a given neutron chemical potential $\mu_n$ (and $T=0$) the free energy depends on the values of the background fields $\phi$, $\omega$, $\rho$, the CDW wavenumber $q$, and the electron chemical potential $\mu_e$. These quantities are computed via the coupled stationarity equations
\be
\frac{\partial\Omega}{\partial \phi} =\frac{\partial\Omega}{\partial \omega} =\frac{\partial\Omega}{\partial \rho} =\frac{\partial\Omega}{\partial q} =\frac{\partial\Omega}{\partial \mu_e}=0  \, . 
\ee
The derivative with respect to $\mu_e$ is the (negative of the)
electric charge density, which we require to vanish (locally).  
The explicit form of the stationarity equations is 
\begin{subequations}\label{stat}
\bea
g_\sigma n_s &=& -\tilde{U}'(\phi)-4q^2\phi\left[1-\frac{g_\sigma^2}{4\pi^2}\left(1+\ln\frac{M^2}{\ell^2}\right)\right]-(1-\delta_{0q})\epsilon+ \frac{g_\sigma q^3}{2\pi^2}F'(y) 
 \, , \label{eq:sigma} \\[2ex]
g_\omega n_B&=&m_\omega^2\omega+\omega(d_\omega\omega^2+d_{\omega\rho}\rho^2)  \, , \label{eq:omega} \\[2ex]
g_\rho n_I&=&m_\rho^2\rho+\rho(d_{\omega\rho}\omega^2+d_\rho \rho^2)  \, , \label{eq:rho} \\[2ex]
j &=& -4q\phi^2\left(1-\frac{g_\sigma^2}{4\pi^2}\ln\frac{M^2}{\ell^2}\right)+\frac{q^3}{2\pi^2}[4F(y)-yF'(y)] \, , \label{eq:q}\\[2ex]
n_p&=&n_e+n_\mu \, .\label{eq:mue}
\eea
\end{subequations}
These equations contain the following abbreviations. The scalar density is
\bea
n_s &=& -\sum_{N=n,p} \frac{\partial{\cal P}(M,q,\mu_N^*)}{\partial M} \, ,
\eea
with ${\cal P}$ from Eq.\ (\ref{calD}), while baryon and isospin densities are 
\begin{subequations} \label{nBI}
\bea
n_B &=& n_n+n_p = \frac{\partial{\cal P}(M,q,\mu_n^*)}{\partial \mu_n} + \frac{\partial{\cal P}(M,q,\mu_p^*)}{\partial \mu_p}\, ,\\[2ex]
n_I &=&n_n-n_p = \frac{\partial{\cal P}(M,q,\mu_n^*)}{\partial \mu_n} - \frac{\partial{\cal P}(M,q,\mu_p^*)}{\partial \mu_p}\, ,
\eea
\end{subequations}
where $n_n$ and $n_p$ are neutron and proton number densities. The nucleonic contribution to the axial current is 
\be
j = -\sum_{N=n,p}\frac{\partial {\cal P}(M,q,\mu_N^*)}{\partial q}
\, , 
\ee
and, finally, the lepton densities are 
\be
n_\ell = \Theta(\mu_\ell-m_\ell)\frac{(\mu_\ell^2-m_\ell^2)^{3/2}}{3\pi^2} \, , \qquad \ell = e,\mu \, .
\ee
Once the stationarity equations are solved, we can insert the results into the free energy density, which gives the pressure $P=-\Omega$. For the calculation of the equation of state, needed later in the construction of the neutron stars, we also have to compute the energy density
\be \label{epsilon}
\varepsilon=  -P + \mu_e n_e+ \mu_\mu n_\mu + \mu_B n_B+\mu_I n_I =  -P + \mu_n n_B \, .
\ee
Due to beta equilibrium and electric charge neutrality the various terms originating from the 
Legendre transform of the pressure collapse to a single term $\mu_n n_B$. This implies that adding {\it any} fermion to the system has an energy cost $\mu_n$. This is obvious for the neutron, which has baryon number $N_B=1$ and chemical potential $\mu_n$. Adding a proton, which also has $N_B=1$, corresponds to a cost  $\mu_n = \mu_p+\mu_\ell$ because one also has to add an electron (or muon) to maintain electric charge neutrality. Similarly, adding an electron (or muon) also requires to add a proton at the same time, resulting in the same energy cost.

\subsection{Parameter fit}
\label{sec:para}

It remains to fit the parameters of the model. Rather than working with a single parameter set, we will explore the parameter space within and slightly beyond the empirical uncertainties. This enables us to identify regions in parameter space where the CDW is preferred and where realistic stars are located and to check whether these two regions overlap. Also, having in mind that our model is of phenomenological nature, we want to allow for some freedom in the parameter choice to identify tendencies or phases that might violate empirical constraints within our model but which could nevertheless be realized in QCD. 

We need to determine 11 yet unknown model parameters: 
the parameters of the vacuum potential $\epsilon,a_1,a_2,a_3,a_4$, the vector meson self-couplings $d_\omega,d_\rho,d_{\omega\rho}$, and the Yukawa couplings $g_\sigma,g_\omega,g_\rho$. We will fit them with the help of vacuum properties, properties of isospin-symmetric nuclear matter at saturation density, and properties of pure neutron matter at half times saturation density. The fit to vacuum properties and to symmetric nuclear matter follows closely the procedure of Refs.\ \cite{Fraga:2022yls,Pitsinigkos:2023xee}, where more details can be found. 
The fit to pure neutron matter is performed here for the first time within the present model. (The binding energy of pure neutron matter was calculated, albeit not used as a fit, in a version of the model without quartic self-interactions of the vector mesons \cite{Drews:2014spa}).  Making use of pure neutron matter is motivated by  the fact that matter under neutron star conditions is highly isospin asymmetric. Hence, reproducing the physics  of pure neutron matter is at least as important as a realistic description of symmetric nuclear matter \cite{Alford:2022bpp}. 

The numerical values of the physical quantities used for the fit are as follows. We use the pion mass $m_\pi=140\, {\rm MeV}$ and the saturation density of isospin-symmetric nuclear matter $n_0=0.153\, {\rm fm}^{-3}$ with corresponding binding energy $E_0=-16.3\, {\rm MeV}$, which implies a chemical potential for the first-order baryon onset of $\mu_0=922.7\, {\rm MeV}$. Moreover, we use the asymmetry energy of nuclear matter at saturation $S = 32\, {\rm MeV}$, and the binding energy of pure neutron matter $E_0=+ 10\, {\rm MeV}$ at $n_B=0.5\,n_0$.  These quantities are all relatively well known such that we may assume that reasonable variations will not change our final conclusions. Less well known quantities are the incompressibility of nuclear matter at saturation, often assumed to be in the range $K\simeq (200-300)\, {\rm MeV}$ \cite{glendenning}, possibly narrowed down by various experimental and theoretical approaches to $K\simeq (200-270)\, {\rm MeV}$ 
\cite{Blaizot:1995zz,Vretenar:2003qm,Youngblood:2004fe,Shlomo:2006ole,Liu:2024qds},   and the effective (Dirac) mass of the nucleon at saturation $M_0 \simeq (0.7-0.8)\,m_N$ \cite{glendenning}.  We will mainly use variations in these two quantities for the study of our parameter space. We shall also compute, but not use for the fit, the ``slope parameter'' $L$, which describes the change of the asymmetry energy as the baryon density is varied. In Appendix \ref{app:KSL} we calculate $K$, $S$, and $L$ within our model, leading to the results (\ref{K1}), (\ref{S1}), (\ref{L1}), which are used in the following.

\subsubsection{Vacuum}
\label{sec:vacuum}

In the vacuum, where there is no CDW, i.e., $q=0$, we require $\phi = \langle \sigma\rangle = f_\pi$. This condition and the expression for the pion mass in our model (obtained by temporarily reinstating pion fluctuations) yield $a_1=m_\pi^2\simeq 1.96\times 10^4\, {\rm MeV}^2$, $\epsilon=f_\pi m_\pi^2\simeq 1.80\times 10^6\, {\rm MeV}^3$. Due to $\phi=f_\pi$ the nucleon mass in the vacuum is 
$m_N = g_\sigma f_\pi$, which we use to fix $g_\sigma \simeq 10.2$. 

We consider two additional vacuum quantities, not included in the fit, but useful to get a more complete picture of the properties of the model. Firstly, the sigma mass, obtained by temporarily including the $\sigma$ fluctuations, 
\be \label{msig}
m_\sigma^2 = \tilde{U}''(f_\pi) = m_\pi^2+f_\pi^2 a_2
 \, . 
 \ee
This mass cannot be compared to the mass of an experimentally verified stable quasiparticle. But we can roughly compare the value to the $f_0(500)$ resonance, and, at the very least, it gives us a stability criterion by checking $m_\sigma^2>0$. Secondly, we keep track of the (un)boundedness of the vacuum potential at large $\phi$. On tree level, this is done by checking the sign of $a_4$, but in the presence of the nucleonic fluctuations there is an additional contribution from the Dirac sea, see Eq.\ (\ref{Utilde}), such that for large $\phi$,
\be \label{a8}
\tilde{U}(\phi) = a_{(8)} \phi^8 + {\cal O}(\phi^6) \, , \qquad a_{(8)} \equiv \frac{1}{96}\left(\frac{a_4}{4}-\frac{m_N^4}{\pi^2f_\pi^8}\right) \, .
\ee
Unboundedness in the $\phi$ direction has been observed previously \cite{Pitsinigkos:2023xee}, also in related models that take into account nucleonic vacuum fluctuations \cite{Skokov:2010sf}.  Indeed, we shall find later that for our  most realistic parameter sets we have $a_{(8)}<0$.  However, in contrast to the unboundedness in the $q$ direction, as discussed above, there seem to be no obvious artifacts resulting from this behavior, and it is conceivable that boundedness is achieved by higher-loop contributions or by simply adding higher-order terms in $\phi$ with the correct sign without changing the physical results qualitatively. 

\subsubsection{Symmetric matter at saturation}
\label{sec:symmetric}

We can express 
the Yukawa coupling constants for the vector mesons in terms of the saturation properties of symmetric nuclear matter,
\begin{subequations}\allowdisplaybreaks \label{gwgr}
\bea
g_{\omega}^2 &=& \frac{m_\omega^2}{2n_0}(\mu_0-\mu_0^*)\left[1+\sqrt{1+\frac{4d_\omega n_0(\mu_0-\mu_0^*)}{m_\omega^4}}\right] \, , \\[2ex]
g_{\rho}^2 &=& \frac{3\pi^2m_\rho^2}{k_F^3}\left(S-\frac{k_F^2}{6\mu_0^*}\right)\left(1+\frac{d_{\omega\rho} \omega_0^2}{m_\rho^2}\right) \, , 
\eea
\end{subequations}
where $\mu_0^*= \sqrt{k_F^2+M_0^2}$ and $k_F = (3\pi^2n_0/2)^{1/3}$, and where $\omega_0$ is the omega condensate at saturation \begin{equation}
\omega_0= \frac{g_{\omega}n_0}{m_\omega^2}f(x_0)\, , 
\end{equation}
with 
\begin{equation}
f(x) \equiv \frac{3}{2x}\frac{1-(\sqrt{1+x^2}-x)^{2/3}}{(\sqrt{1+x^2}-x)^{1/3}} \, , \qquad x_0\equiv \frac{3\sqrt{3d_\omega}\,g_{\omega}n_0}{2m_\omega^3} \, . 
\end{equation}
Using the numerical values of all fixed physical parameters, the relations (\ref{gwgr}) yield functions 
\be \label{gg}
g_\omega = g_\omega(d_\omega,M_0) \, , \qquad  g_\rho = g_\rho(d_\omega,d_{\omega\rho},M_0) \, .
\ee
Next, we couple the expression of the incompressibility $K$ (\ref{K1}), the condition that the pressure at saturation equals the vacuum pressure, and the stationarity equation for  $\phi$ at saturation to derive analytical (but very lengthy) relations 
\be \label{a234}
a_2 = a_2(d_\omega,M_0,K) \, , \qquad a_3 = a_3(d_\omega,M_0,K) \, , \qquad a_4 = a_4(d_\omega,M_0,K) \, .
\ee
Since $\rho=0$ for symmetric nuclear matter, $d_\rho$ and $d_{\omega\rho}$ do not appear in these relations.

\subsubsection{Pure neutron matter}
\label{sec:neutron}

  Besides the freedom in $M_0$ and $K$, it remains to fix the vector meson self-couplings $d_\omega$, $d_\rho$, $d_{\omega\rho}$. 
If the mesonic self-interactions are constructed from chirally invariant terms, one finds that there are two independent couplings, $d_\omega=d_\rho$ and $d_{\omega\rho}$, see for instance Appendix A of Ref.\ \cite{Fraga:2022yls}. Since chiral symmetry is broken explicitly, one may introduce a slight violation of (or, as a more phenomenological approach, completely ignore) this constraint. For simplicity, we will reduce the number of independent couplings to 1 by two different assumptions that do respect this constraint, namely 
\begin{subequations} \label{fit12}
\bea
\mbox{Fit(ddd):} \qquad && d\equiv d_\omega = d_\rho = d_{\omega\rho} \, , \label{fit1} \\[2ex]
\mbox{Fit(00d):} \qquad  &&d_\omega=d_\rho=0 \, , \qquad d\equiv d_{\omega\rho} \neq 0 \, . \label{fit2} 
\eea
\end{subequations}
The main idea behind these two approaches is as follows. Fit(ddd) helps us to connect our results to Ref.\ \cite{Pitsinigkos:2023xee} and enables us to discuss a scenario where the CDW covers a relatively large region of the parameter space. (In Ref.\ \cite{Pitsinigkos:2023xee}, a nonzero $d\equiv d_\omega$ was considered; $d_\rho$ and $d_{\omega\rho}$ were absent since only symmetric nuclear matter and thus $\rho=0$ was considered.)  However, as we will see momentarily, Fit(ddd) is unphysical already in its predictions for pure neutron matter and, as will become clear later,  does not meet basic astrophysical constraints. It is nevertheless useful for the interpretation of our results in that it contrasts the results of Fit(00d), which, as we shall see,  {\it can} reproduce realistic neutron matter and yield realistic neutron stars. 

\begin{figure} [t]
\begin{center}
\includegraphics[width=0.45\textwidth]{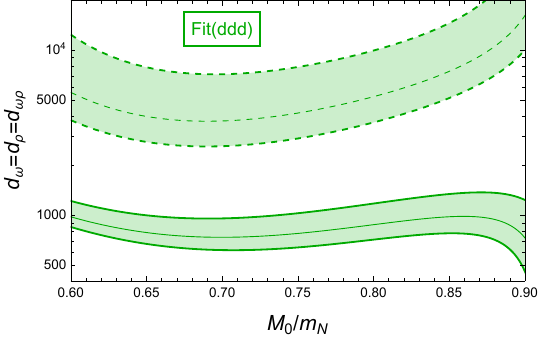}\hspace{0.3cm}\includegraphics[width=0.45\textwidth]{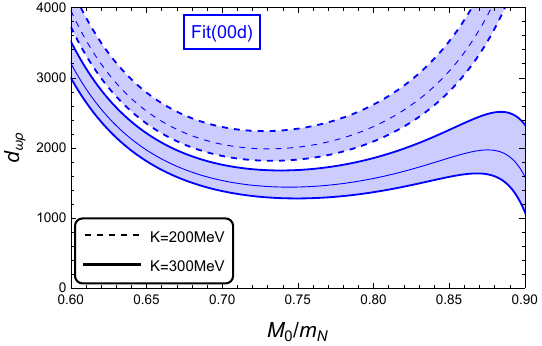}
\caption{Values of the vector meson self-coupling constants following  Fit(ddd) (\ref{fit1}) (left panel) and Fit(00d) (\ref{fit2}) (right panel) as functions of the effective nucleon mass at saturation $M_0$. In both panels, the band enclosed by solid (dashed) lines is for $K=300 \,(200)\, {\rm MeV}$. Each band is obtained by varying the energy per nucleon of pure neutron matter in the range $E_0=[9.6,10.6] \, {\rm MeV}$. The curves in the middle of the bands are given by $E_0=10\, {\rm MeV}$, the value used for all following results of the paper. (Note that the vertical scale is logarithmic in the left panel and linear in the right panel.)}   
\label{fig:E01}
\end{center}
\end{figure}

For our fits we use the binding energy $E_0$ of pure neutron matter from $\chi$EFT \cite{Tews:2018kmu}, also used in Ref.\ \cite{Alford:2022bpp}. Within our model, we compute the 
binding energy from the thermodynamic quantities $\varepsilon$ and $n_B$ via 
\be
E_0 = \frac{\varepsilon}{n_B} - m_N \, . 
\ee
Then, for given $M_0$ and $K$, we solve the condition $E_0=10\, {\rm MeV}$ together with $n_B=0.5\,n_0$ and the three stationarity equations (\ref{eq:sigma}), (\ref{eq:omega}), (\ref{eq:rho}) for $\phi,\omega,\rho,\mu_n,d$, with $\mu_p=\mu_e=0$, using the relations (\ref{gg}) and (\ref{a234}). In Fig.\ \ref{fig:E01} we show the resulting $d=d(M_0,K)$ for the two fits (\ref{fit12}) (left and right panels) as a function of $M_0$ for two different values of $K$ (solid and dashed curves). In this figure we explore the effect of variations in $E_0$ within the width of the error band of $\chi$EFT, while in the rest of the paper we will always work with $E_0=10\, {\rm MeV}$, shown in the middle of the bands of Fig.\ \ref{fig:E01}. To put the values $d\sim 10^3-10^4$ into context it is useful to quantify the effect of the vector meson self-couplings on the condensates. For a simple estimate let us assume isospin symmetry. Then, expanding the solution of the stationarity equation (\ref{eq:omega}) gives
\bea
\omega = \frac{g_\omega n_B}{m_\omega^2}\left[1-d_\omega\frac{g_\omega^2 n_B^2}{m_\omega^6} + {\cal O}(d_\omega^2)\right] \, .
\eea
For a typical value $g_\omega =10$ (see for instance Table \ref{tab:para} in Sec.\ \ref{sec:locate}) and saturation density $n_B=n_0$ we have 
\be
\frac{g_\omega^2 n_0^2}{m_\omega^6} \simeq 6\times 10^{-4} \, .
\ee
 Therefore, values $d_\omega\sim 10^3-10^4$ are ``reasonable'' since they lead to corrections of order 1 to the omega condensate. Similar estimates apply to the rho meson condensate once isospin asymmetry is reinstated. We also see that the effect of the quartic vector meson couplings becomes stronger at larger baryon (and isospin) densities. Different normalizations of the coupling constants are possible and sometimes used in the literature \cite{Steiner:2004fi,Alford:2022bpp}. For instance, $d_\omega/g_\omega^4$, $d_\rho/g_\rho^4$, $d_{\omega\rho}/(g_\omega^2 g_\rho^2)$ are (roughly) of order 1. These rescalings have the additional advantage that the model then only depends on the ratios $g_\omega/m_\omega$, $g_\rho/m_\rho$ rather than $g_\omega$, $g_\rho$ and $m_\omega$, $m_\rho$ separately (if the rescaled condensates $g_\omega\omega$, $g_\rho\rho$ are used). Our convention, on the other hand, is somewhat more straightforward and thus we will keep using $d_\omega$, $d_\rho$, $d_{\omega\rho}$ even though their values seem unnaturally large. 

\begin{figure} [t]
\begin{center}
\hbox{
\includegraphics[width=0.33\textwidth]{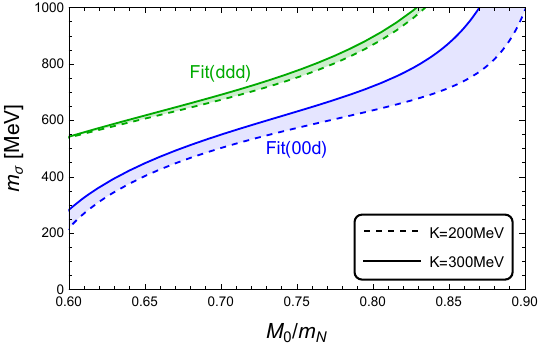}
\includegraphics[width=0.33\textwidth]{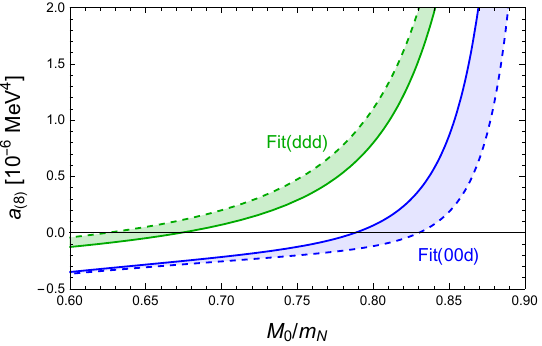}
\includegraphics[width=0.33\textwidth]{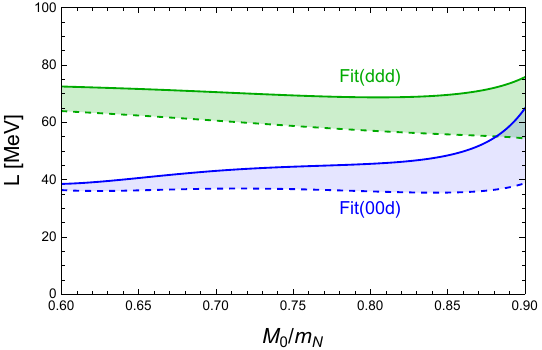}}
\caption{Vacuum quantities (sigma mass $m_\sigma$, left panel, and coefficient $a_{(8)}$, middle panel) and slope parameter of nuclear matter $L$, as  functions of $M_0$ for the two different fits (\ref{fit12}). For each curve, we are using the value $E_0=10\, {\rm MeV}$ at $n_B = 0.5\, n_0$ for pure neutron matter, i.e., the curves in the middle of the bands of Fig.\ \ref{fig:E01}. }
\label{fig:L}
\end{center}
\end{figure}

Once the value of $d$ is fixed, we can go back to Eqs.\ (\ref{gg}) and (\ref{a234}) and determine the Yukawa couplings  $g_\omega$, $g_\rho$ and the parameters of the scalar potential $a_2$, $a_3$, $a_4$ (these parameters vary as we vary $M_0$ while keeping all vacuum and low-density properties fixed). 
Having fixed all parameters we may compute the vacuum quantities $m_\sigma$ (\ref{msig}), $a_{(8)}$ (\ref{a8}), and the slope parameter $L$ (\ref{L1}). The results for the two fits are shown in Fig.\ \ref{fig:L}.
We find that $m_\sigma^2>0$ for all values of $K$ and $M_0$ used here and that agreement with the (very broad) $f_0(500)$ resonance is achieved for intermediate and small values of $M_0$ for Fit(00d) and Fit(ddd), respectively. 
The middle panel shows that the vacuum potential is unbounded in the $\phi$ direction in a 
large range of small and intermediate values of $M_0$ for Fit(00d) and for small values of $M_0$ for Fit(ddd). 
Interestingly, for neither fit does the slope parameter vary much with $M_0$. Empirical studies and ab initio calculations currently  still leave a very large uncertainty in $L$
\cite{PREX:2021umo,Hu:2021trw,Sotani:2022hhq,CREX:2022kgg,Lattimer:2023rpe,Giacalone:2023cet}. We shall come back to our prediction for $L$ once we have narrowed down the allowed parameter region with the help of neutron star constraints; see Sec.\ \ref{sec:locate} and in particular Table \ref{tab:para}.

\begin{figure} [t]
\begin{center}
\includegraphics[width=0.48\textwidth]{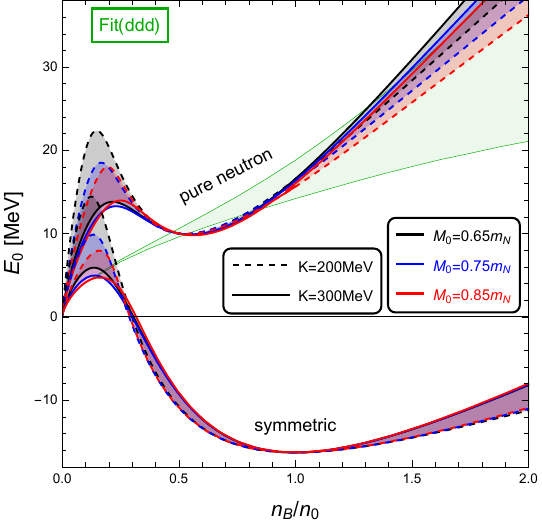}\hspace{0.5cm}
\includegraphics[width=0.48\textwidth]{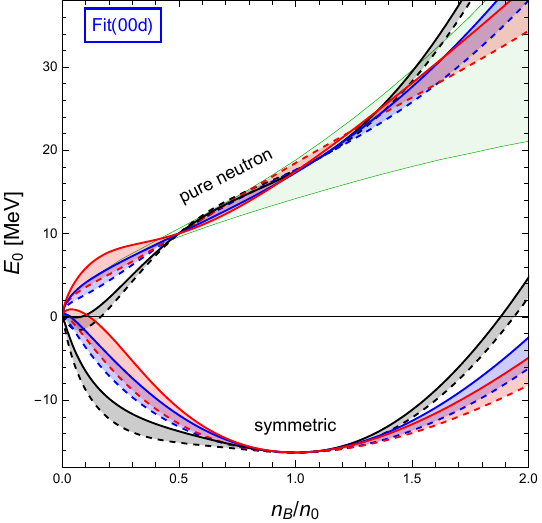}
\caption{Energy per nucleon relative to the vacuum mass as a function of baryon density for the two different fits (\ref{fit12}) (left and right panels). Each panel shows the results for pure neutron matter and isospin-symmetric nuclear matter, for three different values of $M_0$ (black, blue, red) and two different values of $K$ (solid, dashed). For comparison, the green band is the pure neutron matter result from $\chi$EFT \cite{Tews:2018kmu}. Both fits are constructed to reproduce the points $(n_B,E_0)=(n_0,-16.3\, {\rm MeV})$ for symmetric nuclear matter and $(n_B,E_0)=(0.5\,n_0,+10\, {\rm MeV})$ for pure neutron matter. }
\label{fig:E02}
\end{center}
\end{figure}

The density $n_B=0.5\,n_0$ at which we reproduce the binding energy is of course somewhat arbitrary. Therefore, it is useful to compare our model predictions for $E_0$ to $\chi$EFT over a wider range in density. This is done in Fig.\ \ref{fig:E02} for both fits (\ref{fit12}). In each case, we show the results for three different values of $M_0$ and for each $M_0$ we use two values of the incompressibility, $K=(200,300)\, {\rm MeV}$, resulting in the 
colored bands. The bands are shown in comparison to the error band from $\chi$EFT, taken from Ref.\ \cite{Alford:2022bpp}, originally computed in Ref.\ \cite{Tews:2018kmu}.
For completeness, we have also included the corresponding results of our model for symmetric nuclear matter. 

We see in the left panel for Fit(ddd) that using the single value of $E_0$ at $n_B=0.5\,n_0$ can lead to strong deviations from $\chi$EFT at low densities. In particular, we see a strong non-monotonicity, which is an indication of a first-order phase transition. In symmetric nuclear matter, we know that such a phase transition exists, namely the baryon onset at $\mu_B=\mu_0$ from $n_B=0$ to $n_B=n_0$. The results of $\chi$EFT suggest that this discontinuity disappears as matter is made more and more isospin asymmetric. We shall see later that for neutron star matter (which is somewhere between symmetric matter and pure neutron matter), a first-order transition within nuclear matter indeed exists within Fit(ddd) for any realistic $M_0$, while with Fit(00d) there are parameter regions without such a transition.

The left panel of Fig.\ \ref{fig:E02} suggests that additionally constraining the slope of the function $E_0(n_B)$ might be necessary to
make the fit more realistic. We have checked that this is possible for instance by allowing for independent and nonzero couplings $d_\omega=d_\rho$ and $d_{\omega\rho}$. However, we see from the right panel that Fit(00d) produces ``automatically'' the right slope for certain values of $M_0$ and shows excellent agreement with $\chi$EFT: for instance, the blue solid curve for $M_0=0.75\,m_N$ and $K=300\, {\rm MeV}$ lies within the $\chi$EFT band for all densities shown here. We shall therefore stick to the two fits (\ref{fit12}) and leave any further 
exploration of the freedom from the three coupling constants $d_\omega$, $d_\rho$, $d_{\omega\rho}$ to future studies.

\section{Results}
\label{sec:results}

For fixed model parameters and a given thermodynamic parameter $\mu_n$, we can now solve the stationarity equations (\ref{stat}) numerically, compute the corresponding free energy density (\ref{OmPmat}), and thus determine the branch of the solution that is energetically preferred for each $\mu_n$.  For definiteness, we restrict the parameter space by fixing $K=300\, {\rm MeV}$ in all following results, but 
we keep varying the parameter $M_0$ to explore qualitatively different scenarios. 
 The choice of $K$, at the upper end of the allowed range, perhaps even somewhat beyond that range, is motivated as follows. As we shall see, the most stringent constraint for the CDW will arise from the neutron star masses. With nuclear matter as incompressible as possible we expect to obtain the largest possible masses. If we found realistic neutron stars with CDW cores in this approach, this would not allow us to draw firm conclusions about the astrophysical presence of CDW matter because realistic parameter sets will likely lead to smaller maximum neutron star masses, possibly violating known constraints. However, as we shall find no CDW core -- even for the most (perhaps unrealistically) incompressible nuclear matter -- we will be able to reach a firm conclusion. Hence, for the main purpose of this paper, there is no need to discuss results for smaller incompressibilities 
and we will keep $K=300\, {\rm MeV}$ fixed in the following. (We have checked that indeed $K=200\, {\rm MeV}$ does give smaller maximal neutron star masses for all $M_0$.) 

\subsection{Phase diagram}
\label{sec:phase}

\begin{figure} [t]
\begin{center}
\hbox{\includegraphics[width=0.5\textwidth]{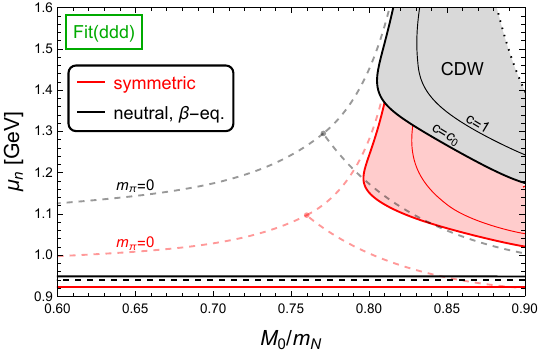}\includegraphics[width=0.5\textwidth]{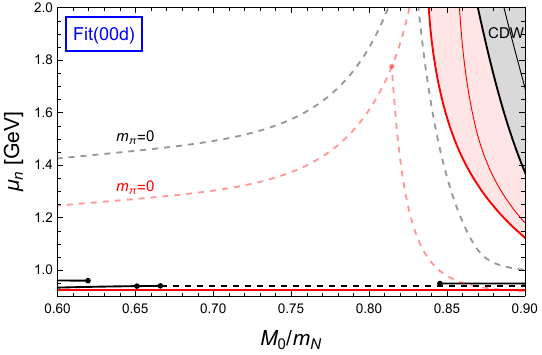}}
\caption{Phase structure in the plane of the neutron chemical potential $\mu_n$ and the model parameter $M_0$ (as $M_0$ is varied, the other model parameters are adjusted to keep the vacuum and low-density physics fixed). The figure compares the two fits (\ref{fit12}) (left and right panels) and the results for symmetric nuclear matter (red) with electrically neutral, beta-equilibrated matter (black). Additionally, the chiral and CDW phase transitions for the chiral limit $m_\pi=0$ are shown (pale dashed curves). Solid (dashed) curves represent first-order (second-order) phase transitions. The CDW (at physical pion mass) is energetically favored in the shaded area for renormalization scale parameter $c=c_0\simeq 0.3$. The thin solid line within the shaded area would be the transition for $c=1$, used in Ref.\ \cite{Pitsinigkos:2023xee}. In the left panel, for $c=c_0$, the CDW has an unphysical upper boundary (dotted line), reminiscent of the artifacts encountered for $c=0$ in previous literature.}
\label{fig:muM01}
\end{center}
\end{figure}

We start with the phase structure for the two fits Fit(ddd) and Fit(00d), see left and right panels of Fig.\ \ref{fig:muM01}. This figure combines the results for isospin-symmetric matter and matter under neutron star conditions as well as the chiral limit and the case of a physical pion mass. Let us first comment on the chiral and CDW transitions, which show qualitatively the same behavior 
in symmetric matter and  neutron star matter and in both fits: In the chiral limit, $m_\pi=0$, the model shows a second-order chiral phase transition. For small $M_0$, this is a transition from isotropic nuclear matter to an isotropic phase with massless fermions. For large $M_0$, isotropic nuclear matter is superseded by a CDW in a second-order ``large wavelength'' transition, where the CDW wavenumber $q$ increases continuously from zero, before chiral symmetry is restored in a second-order ``small amplitude'' transition, where the magnitude of the chiral condensate $\phi$ goes to zero continuously at nonzero wavenumber $q$. In the presence of the physical pion mass, the isotropic chiral transition at small $M_0$ becomes a smooth crossover. This is a consequence of taking into account the fermionic vacuum fluctuations, as pointed out in Ref.\ \cite{Pitsinigkos:2023xee}. In contrast, in the ``no-sea approximation'', the chiral transition would be of first order for all relevant choices of the model parameters. The anisotropic phase at large $M_0$ is now bounded by a discontinuous transition. Moreover, the range in $\mu_n$ where the CDW is energetically favored is reduced compared to the chiral limit, i.e., switching on the pion mass disfavors the CDW. 

The red curves with Fit(ddd) in the left panel are similar, although not identical, to the results of Ref.\ \cite{Pitsinigkos:2023xee}. The differences are that, firstly, in Ref.\
\cite{Pitsinigkos:2023xee} $d_\omega$ was not adjusted for each $M_0$ to reproduce properties of pure neutron matter. Secondly, in Ref.\ \cite{Pitsinigkos:2023xee} the renormalization scale (\ref{ell}) was used with $c=1$. In Fig.\ \ref{fig:muM01} we show the result for $c=1$ (thin solid curves) in comparison to $c=c_0\simeq 0.3$ with $c_0$ from Eq.\ (\ref{c0}) (thick solid boundaries of the shaded CDW area). 
Using different values of $c$ gives us an idea about the sensitivity of the result with respect to variations of the renormalization scale. Moreover, the value $c=c_0$ corresponds to the most favorable choice for the CDW. More precisely, it extends the CDW down to the lowest possible values of $\mu_n$ without making the 
effective potential unbounded in the $q$ direction. Therefore, for each $M_0$, we are setting a lower limit in $\mu_n$ above which the CDW can be favored, which will be useful for our main conclusions.  The price we have to pay for this choice is that unphysical features in the free energy -- observed with $c=0$ in previous models, resolved with $c=1$ in Ref.\ \cite{Pitsinigkos:2023xee} -- reappear with $c=c_0$ at large chemical potentials (and large $M_0$). Namely, at a certain chemical potential, the CDW solution ceases to be a minimum of the effective potential, see inset in Fig.\ 5 of Ref.\ \cite{Pitsinigkos:2023xee}. We have indicated this upper, unphysical, boundary of the CDW region by a dotted line in the left panel of Fig.\ \ref{fig:muM01}. (This line, while clearly visible for neutral, beta-equilibrated matter, also exists for symmetric nuclear matter, barely visible in the figure.) In the right panel, using the more realistic Fit(00d), the unphysical boundary exists as well but is beyond the scale shown here. 

Besides the chiral and CDW transitions, Fig.\ \ref{fig:muM01} also shows the onset of nuclear matter and various phase transitions at low densities. In symmetric nuclear matter there is, by construction, for any $M_0$, a first-order baryon onset at $\mu_n=\mu_B = \mu_0$ (horizontal red solid line). In the left panel, with Fit(ddd), we see that the onset of neutral, beta-equilibrated matter 
is of second order, at $\mu_n=m_N$ (horizontal dashed black line), before there is a first-order phase transition within nuclear matter (almost horizontal solid black curve). This was already anticipated from the (unphysical) non-monotonic binding energy of pure neutron matter, see left panel of Fig.\ \ref{fig:E02}. Using Fit(00d), on the other hand, the low-density first-order transition has disappeared for intermediate -- most realistic -- values of $M_0$, consistent with the monotonic behavior of the binding energy of neutron matter in the right panel of Fig.\ \ref{fig:E02}. For $M_0\lesssim 0.65\, m_N$ we encounter  a first-order onset of neutral, beta-equilibrated matter and for even smaller values, $M_0\lesssim 0.62$ there is a first-order onset {\it and} another first-order transition within nuclear matter. For $0.65\, m_N\lesssim M_0\lesssim 0.67\, m_N$
and $M_0\gtrsim 0.85\, m_N$ there is a second-order onset  followed by a first-order transition, the same scenario as for the entire range shown for Fit(ddd). 

The most important conclusion of Fig.\ \ref{fig:muM01} concerns
the effect of charge neutrality and beta-equilibrium on the phase diagram. This effect is significant already in the simplest scenario of $m_\pi=0$ and small $M_0$, i.e., an isotropic chiral phase transition. For instance with Fit(ddd) and $M_0=0.75\, m_N$, the critical chemical potential of the chiral transition increases from $\mu_n\simeq 1.08\, {\rm GeV}$ for symmetric matter to $\mu_n\simeq 1.24\, {\rm GeV}$ for matter under neutron star conditions. The value of $M_0$ beyond which there is a CDW region, however, does not change much. For instance, in the  chiral limit, and within Fit(ddd), this critical $M_0$ increases  from $M_0\simeq 0.76\, m_N$ to $M_0\simeq 0.77\, m_N$. The same qualitative behavior can be observed for Fit(00d). The difference between the red and black shaded areas in both panels show that  these trends persist in the presence of a physical pion mass; in particular, the neutron star conditions postpone the first-order onset of the CDW to larger chemical potentials. We thus conclude, even before constructing any neutron star explicitly, that the CDW is disfavored by the conditions present in the star.

\subsection{Mass-radius curves}
\label{sec:MR}

We construct neutron stars from our model by solving the Tolman-Oppenheimer-Volkoff (TOV) equations \cite{tolman,PhysRev.55.364,PhysRev.55.374}
\begin{subequations} \label{TOVs}
\bea
\frac{\partial P}{\partial r} &=& -\frac{G}{r^2} \frac{(M+4\pi Pr^3)(\varepsilon+P)}{1-\frac{2GM}{r}} \, , \label{TOV1}\\[2ex]
 \frac{\partial M}{\partial r} &=& 4\pi r^2 \varepsilon \, , \label{TOV2}
\eea
\end{subequations}
where $G=6.709 \times 10^{-39}\, {\rm GeV}^{-2}$ is the gravitational constant, and $P(r)$, $M(r)$, $\varepsilon(r)$ are pressure, mass, and energy density, respectively, as a function of the radial coordinate $r$. From our nucleon-meson model we compute $P$ and $\varepsilon$ at zero temperature, which yields 
the equation of state $P(\varepsilon)$, including discontinuities due to the various possible phase transitions shown in Fig.\ \ref{fig:muM01}. Then, for a given central pressure $P(r=0)=P_0$ as well as $M(r=0)=0$ as boundary conditions we solve the TOV equations numerically. The radius $R$ of the star is identified from the solution by $P(r=R)=0$ because 
the vacuum pressure is zero and the total mass of the star is then given by  $M\equiv M(R)$. Repeating this procedure for many $P_0$ yields the mass-radius curve for a given equation of state, i.e., for a given parameter set of our model. 

For simplicity, we do not attempt to construct a crystalline phase from our model at low densities that would constitute the crust of the star. Nor do we use a separate equation of state for the crust from the literature. Such a separate equation of state could be connected to our result by assuming a critical chemical potential for the 
crust-core transition. For our purpose, the lack of a realistic crust is irrelevant. The reason is that the maximal mass of the star is essentially independent of the properties of the crust, see for instance Ref.\ \cite{Kovensky:2021kzl} (while the radius does depend strongly on the crust). And it is the maximal mass alone that will turn out to be sufficient to constrain the stars with a potential CDW interior. 

Another simplification concerns the assumption of isotropy in the TOV equations (\ref{TOVs}). The CDW breaks rotational symmetry and thus induces longitudinal and transverse components $T_\ell$ and $T_\perp$ in the stress-energy tensor $T^{\mu\nu} = {\rm diag}(T^{00},T_\perp,T_\perp,T_\ell)$ with respect to the wave vector $\vec{q}$ (here aligned with the spatial 3-direction). As a consequence, for example, the speed of sound depends on the angle between the direction of propagation and $\vec{q}$. There are two reasons that justify working in the isotropic limit $T_\ell\simeq T_\perp\simeq P$ for the purpose of solving the TOV equations.  Firstly, it is not obvious why $\vec{q}$ should point into a globally fixed direction in a macroscopic object like a neutron star core. This global alignment is conceivable, especially if there is a globally preferred direction, for example by a sufficiently strong magnetic field (and a time evolution of the star that allows for the alignment). Otherwise, it is natural to assume that there are  microscopic or mesoscopic domains with wave vectors randomly oriented relative to the other domains, similar to the Weiss domains in a ferromagnet. In this case, one can expect the anisotropies to average out on the scale of the star, such that the isotropic calculation is a very good approximation. Secondly, even if a global alignment occurs, the effect on the macroscopic properties of the star depend on the size of the CDW region. As we shall see, our model does not allow for neutron stars with a sizable CDW core, and thus for all cases we consider we expect the isotropic approximation to be valid. 

\begin{figure} [t]
\begin{center}
\includegraphics[width=0.45\textwidth]{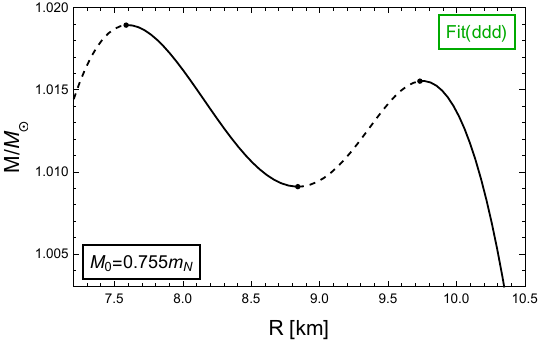}\hspace{0.3cm}
\includegraphics[width=0.45\textwidth]{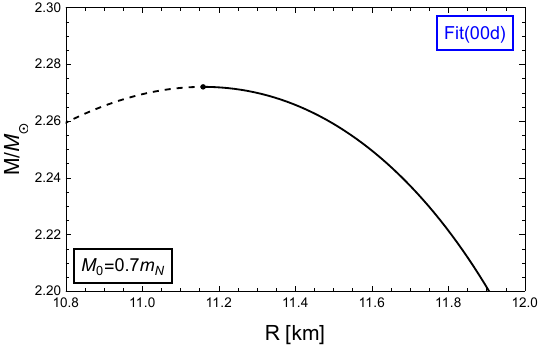}

\includegraphics[width=0.45\textwidth]{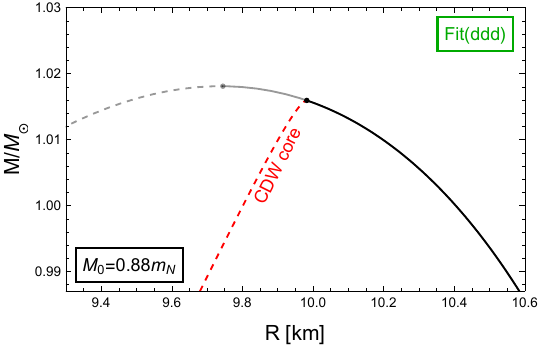}\hspace{0.3cm}
\includegraphics[width=0.45\textwidth]{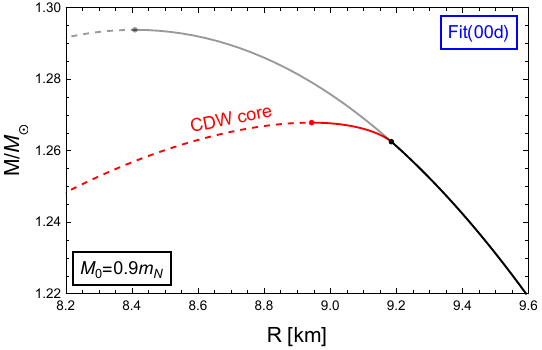}
\caption{Mass-radius curves for neutron stars using Fit(ddd) (\ref{fit1}) (left panels) and Fit(00d) (\ref{fit2}) (right panels), for different values of $M_0$ and fixed $K=300\, {\rm MeV}$. The masses are given in units of the solar mass $M_\odot$. The panels illustrate twin star solutions (upper left), the case of a realistic maximal mass (upper right), unstable stars with a CDW core (lower left), and stars with a stable CDW core (lower right). CDW cores only occur in unrealistic scenarios with maximal masses far below masses of the most massive known neutron stars.}
\label{fig:MR}
\end{center}
\end{figure}

We show four mass-radius curves in Fig.\ \ref{fig:MR}, all obtained with the physical pion mass and with parameters given in the labels and caption of the figure. These four cases are selected to show qualitatively distinct scenarios that are important for the more systematic discussion in the next subsection. The upper left panel shows the case of a twin star configuration, which allows for two stable stars with the same mass and different radii. Here and in the three other panels, dashed segments of the curves indicate unstable stars with respect to radial oscillations \cite{1966ApJ...145..505B,glendenning}.  The upper right panel is the only one in this figure where the maximal mass of the star is larger than two solar masses, thus fulfilling a necessary astrophysical constraint. The lower two panels show two different scenarios containing stars with a CDW core. In the left panel, the stars become unstable with respect to radial oscillations immediately when a CDW core develops, while the right panel shows a case where there is a (small) segment with stable CDW stars. Both panels also show, for comparison, the corresponding curve obtained by ignoring the CDW (grey segments), i.e., assuming matter to be isotropic even though the CDW has lower free energy.

\subsection{Locating realistic neutron stars in the phase diagram}
\label{sec:locate}

\begin{figure} [t]
\begin{center}
\includegraphics[width=0.5\textwidth]{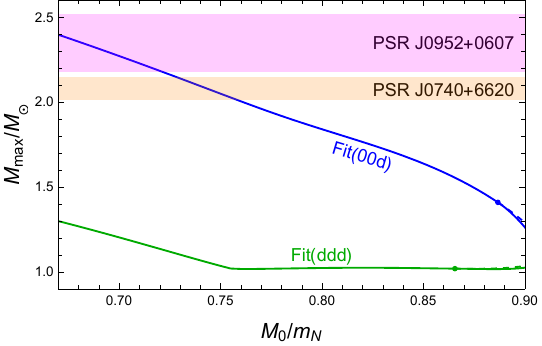}
\caption{Maximal neutron star mass for the two fits used in Fig.\ \ref{fig:muM01} as a function of the Dirac mass at saturation $M_0$. In both curves, only to the right of the dots, CDW cores are possible. The dashed lines (barely distinguishable from the solid lines) are obtained by ignoring the CDW. The solid continuation of the curves represent maximum mass stars with CDW core (blue) and without CDW core -- yet different from the dashed segment because the CDW makes the heaviest stars unstable (green).  The masses of the two most massive known stars PSR J0740+6620 and PSR J0952+0607 including uncertainty bands are taken from Refs.\ \cite{Fonseca:2021wxt} and \cite{Romani:2022jhd}. }
\label{fig:Mmax}
\end{center}
\end{figure}

\begin{figure} [t]
\begin{center}
\includegraphics[width=0.45\textwidth]{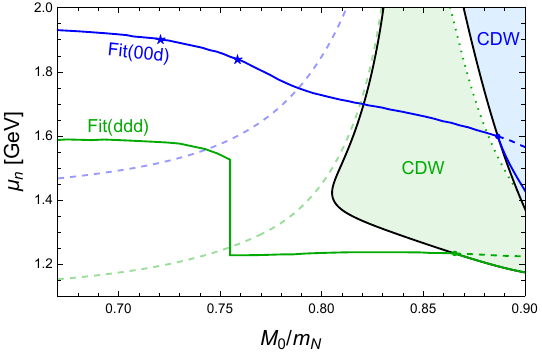}\hspace{0.3cm}
\includegraphics[width=0.45\textwidth]{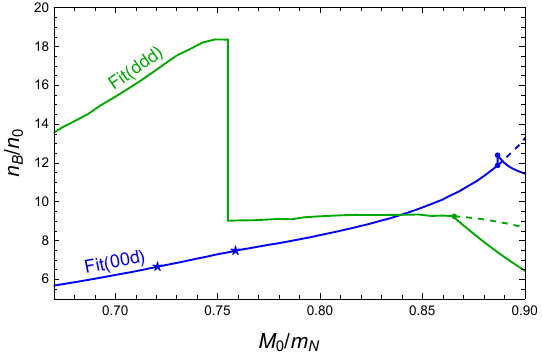}
\caption{Solid blue and green lines show neutron chemical potential (left) and baryon density (right) in the center of the most massive stars of Fig.\ \ref{fig:Mmax}, as a function of the Dirac mass at saturation $M_0$. Dashed continuations of these curves are obtained by ignoring the CDW. The CDW regions (shaded areas) and the chiral phase transition curves in the chiral limit (pale dashed lines) in the left panel are taken from Fig.\ \ref{fig:muM01}. Only the segments on the blue curves to the left of the star markers meet the astrophysical constraint from PSR J0740+6620 ($M_{\rm max}\gtrsim 2.01\, M_\odot$, right marker) and additionally from PSR J0952+0607 ($M_{\rm max}\gtrsim 2.18 \, M_\odot$, left marker). }
\label{fig:Mumax}
\end{center}
\end{figure}

For a systematic study of whether there is a CDW in the core of neutron stars we solve the TOV equations for many values of  $M_0$ (and the resulting sets of model parameters), covering the range shown in Fig.\ \ref{fig:muM01}, and for both Fit(ddd) and Fit(00d). We show the resulting maximal mass $M_{\rm max}$ of the neutron star as a function of $M_0$ in Fig.\ \ref{fig:Mmax} and the corresponding neutron chemical potential $\mu_n$ and baryon density $n_B$ in the center of the most massive star in Fig.\ \ref{fig:Mumax}. The left panel of Fig.\ \ref{fig:Mumax}, where the CDW regions are taken from Fig.\ \ref{fig:muM01}, locates neutron stars in the phase diagram: All stable neutron stars ``live'' below the solid green and blue curves [for Fit(ddd) and Fit(00d), respectively], probing a range in $\mu_n$ throughout the profile of the star with a maximal $\mu_n$ below the solid curves (on the curve for the maximally massive star). The main observations of Figs.\ \ref{fig:Mmax} and \ref{fig:Mumax} are as follows.

\begin{itemize}

\item {\bf Stable and unstable stars with CDW core.} Stars with a CDW core only exist for very large $M_0$ for both fits. For Fit(ddd) they are all unstable with respect to radial oscillations. In this case, the mass-radius curves with CDW stars have the shape shown in the lower left panel of Fig.\ \ref{fig:MR}. As a consequence, the maximal $\mu_n$ follows exactly (within the numerical uncertainty) the curve for the first-order CDW onset, see green curve in the left panel of Fig.\ \ref{fig:Mumax}. Using Fit(00d), stable stars with a CDW core are possible and the curve for the maximal $\mu_n$ has a (small) segment inside the CDW region. This  leads to a discontinuity in the maximal $n_B$, see right panel of Fig.\ \ref{fig:Mumax}: Since the onset of the CDW is of first order, even an infinitesimally small CDW core has a different density by a finite amount compared to the maximally massive star without CDW and essentially the same total mass. An example for a parameter set that allows for stable stars with CDW core is given in the lowest row of Table \ref{tab:para}.

\item {\bf Absence of CDW in realistic neutron stars.} All stars with a CDW core are part of mass-radius curves that have maximum masses $M_{\rm max}$ far below the largest observed masses. This is obvious in Fig.\ \ref{fig:Mmax}.

\item {\bf Realistic neutron stars.} Our model does yield realistic maximal neutron star masses  (without CDW core) for Fit(00d). Parameter sets with $M_0\lesssim 0.76\, m_N$ meet the constraint from PSR J0740+6620 while parameter sets with $M_0\lesssim 0.72\, m_N$ meet the constraints from both PSR J0740+6620 and PSR J0952+0607, see upper right panel of Fig.\ \ref{fig:MR} for a corresponding mass-radius curve. We give the explicit values of the model parameters for two cases with $M_{\rm max}>2\, M_\odot$ in the upper two rows of Table \ref{tab:para}. Identifying the realistic parameter regime also allows us to make a prediction for the slope parameter $L$; using the values of the table we find $L\sim (43-45)\, {\rm MeV}$. Allowing for lower values of $M_0$ would change the lower boundary somewhat, but not by much, see also Fig.\ \ref{fig:L} (and lower values of $M_0$ tend to make pure neutron matter unphysical, see right panel of Fig.\ \ref{fig:E02}). Allowing for larger values of $M_0$ is not possible without violating astrophysical constraints.

\item {\bf Hybrid stars.} The results of Fit(ddd) show a discontinuity in the maximal $\mu_n$ and $n_B$, Fig.\ \ref{fig:Mumax}, and a cusp in the curve for  the maximal mass itself, Fig.\ \ref{fig:Mmax}. The reason is the twin star configuration shown in the upper left panel of Fig.\ \ref{fig:MR}: If the critical $M_0$ of the discontinuity is approached from above, the maximal mass star corresponds to the local maximum at {\it larger} radius (smaller central pressure, density, and chemical potential), while approaching it from below, the maximal mass star is the local maximum at {\it smaller} radius (larger central pressure). Twin star configurations are known to appear due to first-order quark-hadron transitions \cite{Christian:2017jni}. Interestingly, we find that they can also originate from a (sufficiently sharp but) smooth crossover. The curves from Fit(00d) are smooth but, in particular for $\mu_n$, we see a shoulder-like structure reminiscent of the discontinuity. To interpret this structure we have added the curves for the chiral phase transition in the chiral limit in the left panel of Fig.\ \ref{fig:Mumax}. Although they can only be a rough indicator for the (pseudo-)critical chemical potential in the case of a physical pion mass, we see that they intersect the $\mu_n$ curves at the discontinuity for Fit(ddd) and (slightly to the right of) its remnant structure for Fit(00d). This suggests that to the left of this structure [$M_0\lesssim 0.755\, m_N$ for Fit(ddd), $M_0\lesssim 0.77\, m_N$ for Fit(00d)] we may interpret our stars as hybrid stars with a chirally restored core. In particular, the heaviest stars for all realistic parameter sets from Fit(00d) are hybrid (and without twin star structure).

\end{itemize}

\begin{table}[t]
\begin{tabular}{||c | c | c | c |c |c|| c | c |c|c|c||} 
 \hline
   $g_{\omega}$ & $g_{\rho}$ 
   &  $a_2$ 
   & $a_3 [{\rm MeV}^{-2}]$ 
   & $a_4 [{\rm MeV}^{-4}]$  & $d$ 
   & $\;$$M_0/m_N$$\;$  & $\;$$m_\sigma\,[{\rm MeV}]$$\;$  & $\;$$K\,[{\rm MeV}]$$\;$ 
   & $\;$$L \,[{\rm MeV}]$$\;$ & $\;$$M_{\rm max}/M_\odot$$\;$ \\ [0.5ex] 
 \hline\hline
  $\;$ 9.22 $\;$ & $\;$5.60$\;$ & $\;$$47.2$$\;$ & $\;$$-2.32\times 10^{-2}$$\;$ & $\;$$ 2.20\times 10^{-5}$$\;$ & $\;$$1.45\times 10^3$$\;$& 0.760 & $\;$647$\;$ & $\;$300$\;$ & 44.6 & 2.01 \\
  \hline
  $\;$ 10.6 $\;$ & $\;$5.86$\;$ & $\;$$33.3$$\;$ & $-4.15\times 10^{-2}$ & $-2.21\times 10^{-5}$ & $1.53\times 10^3$& 0.700 & 549 & 300 & 43.0 & 2.27 \\
 \hline\hline
  $\;$ 4.98 $\;$ & $\;$5.06$\;$ & $\;$$181$$\;$ & $0.718$ & $2.28\times 10^{-3}$ & $1.87\times 10^3$& 0.890 & 1240 & 300 & 58.6 & 1.38 \\
 \hline
\end{tabular}
\caption{Three parameter sets (first 6 columns) with corresponding physical properties (last 5 columns). The upper two rows are representative for the parameter region  where neutron star masses larger than two solar masses are reached (and where there is no CDW). The lower row defines a parameter set where the CDW is favored at large densities but only small neutron star masses are possible. All sets are based on Fit(00d), i.e., the vector meson couplings are $d_\omega=d_\rho=0$, $d\equiv d_{\omega\rho}$. The model parameters not listed here, together with the other physical properties of low-density nuclear matter used for the parameter fit, are fixed throughout the paper as explained in Sec.\ \ref{sec:para}.}
\label{tab:para}
\end{table}

\subsection{Discussion}
\label{sec:discussion}

Our main observation is the absence of the CDW in realistic neutron stars. Let us discuss and interpret this result. We see from Fit(ddd) that it is possible to extend the CDW region further towards more realistic values of $M_0$, but that at the same time the equation of state becomes extremely soft such that realistic neutron stars are out of reach. In contrast, Fit(00d) allows for sufficiently stiff equations of state but the CDW region is restricted to very large values of $M_0$. And even then, the stars with CDW core are not massive enough to meet the astrophysical constraints. The effect of the vector meson self-couplings on the stiffness of the equation of state is easy to identify: It is known that with a nonzero $d_\omega$ the speed of sound squared asymptotes to 1/3 for large densities, while for vanishing $d_\omega$ it asymptotes to 1 \cite{Fraga:2022yls}. This explains the relative softness of all equations of state within Fit(ddd) (additionally, as we have pointed out, this fit does not reproduce known properties of pure neutron matter).  The CDW itself appears to make the equation of state softer, as the lower panels of Fig.\ \ref{fig:MR} indicates: The segments of the mass-radius curves with CDW core are below the ones where the CDW is ignored. This is, however, only one of the reasons for the absence of the CDW in neutron stars. The more important reason is that  the CDW is preferred only for model parameters which predict soft equations of state already in the absence of a CDW. 

One might ask whether our study has excluded the CDW for all possible realistic parameter sets within the given model. Obviously, due to the multi-dimensional parameter space an exhaustive analysis is very difficult. But, having the eventual result in mind, our strategy regarding the parameter fits was directed towards a general statement about the absence of the CDW. We have checked parameter regions that are already as 
favorable as possible for the CDW: Allowing for very large $M_0$ (thereby stretching the empirically allowed range for $M_0$), choosing the most favorable renormalization scale, and making the equation of state as stiff as possible by choosing $K$ at the upper end of the allowed range. The fact that even in this scenario we have found no realistic star with a CDW core -- and did not even come  close -- makes us confident in concluding that in the given model and the given approximations this is a general statement.

We should emphasize that the inclusion of the fermionic vacuum fluctuations plays a crucial role for this result. The Dirac sea contribution not only turns the first-order chiral transition into a smooth crossover in the absence of a CDW, it also drastically reduces the parameter space where the CDW is favored over the isotropic solution. This was discussed in detail in Ref.\ \cite{Pitsinigkos:2023xee} for isospin-symmetric nuclear matter within the same model as discussed here, see in particular 
the right panel of Fig.\ 5 in this reference. In the ``no-sea approximation'', the CDW exists for all values of $M_0$ and at smaller chemical potentials than found here. This suggests that realistic neutron stars with a CDW core can be constructed in the no-sea approximation, which, in fact, is often used in similar models, at least for isotropic phases of nuclear matter.
However, in that approximation, the CDW is ``too favorable'' because in large parts of the parameter space it sets in as soon as baryonic states are populated, unphysically rendering nuclear matter at saturation anisotropic \cite{Pitsinigkos:2023xee}. Therefore, and on theoretical grounds, it seems obvious that the Dirac sea contribution has to be included, rather than simply be dropped without justification. And it seems, therefore, that our conclusion is robust. However, we should keep in mind that our Yukawa coupling constants are not small ($g_\sigma$ of the order of 10) and the mean-field approximation is a drastic simplification to begin with. Therefore,  a controlled approximation would require more sophisticated (strong-coupling) methods  beyond our one-loop calculation of the pressure. More concretely, while at 
weak coupling the Dirac sea contribution is suppressed, it is conceivable that our approximation overestimates the effect of the nucleonic vacuum fluctuations due to the large values of the couplings. And, based on the connection to the CDW just explained, this may result in an underestimate of the importance of the CDW. In this sense, our result should be taken with some care and, ideally, a calculation strictly valid at strong coupling, should be performed to test our prediction.

\section{Summary and outlook}
\label{sec:summary}

We have investigated the fate of a certain anisotropic configuration of nuclear matter, the chiral density wave (CDW), under neutron star conditions. To this end, we have employed a nucleon-meson model that takes into account vacuum fluctuations of the nucleons and that allows for a chiral phase transition. Besides accounting for electric charge neutrality and electroweak equilibrium we have also improved a previous version of the model by using the freedom in the self-coupling constants of the vector mesons to reproduce properties of pure neutron matter at low densities. 

We have explored the parameter space of the model by exploiting empirical and theoretical uncertainties in low-density properties of isospin-symmetric and pure neutron matter. In particular, we have presented our main results as functions of the effective nucleon mass at saturation density (with all model parameters adjusted as we vary this effective mass such that other low-density properties remain fixed and physical). There are regions in parameter space where there is no solution for the CDW configuration, while in other regions -- for large effective nucleon mass at saturation -- the CDW becomes energetically preferred for sufficiently large chemical potentials. We have found that, where the CDW exists, neutron star conditions shift its first-order onset to higher chemical potentials compared to isospin-symmetric nuclear matter. This is one effect of neutron star conditions, disfavoring the CDW.

By combining our microscopic results with gravity via the 
TOV equations, we have constructed neutron stars for many equations of state, each defined by a set of model parameters. Some of these stars contain a CDW core. In a certain parameter regime, the stars with CDW are unstable with respect to radial oscillations, while for other parameter choices there is a stable branch of stars with CDW. For both scenarios, however, the maximal masses of the stars are far below  2 solar masses, failing a necessary criterion for the validity of the parameter set of the model. The model does allow for mass-radius curves with realistic maximal masses; however, only for model parameters that do not allow for a (thermodynamically stable) CDW.  The main conclusion is thus that the CDW is only preferred in a parameter region where the equation of state is too soft, even in the absence of a CDW.

Our study can be improved and extended in the future in several ways, to further understand and validate our predictions. First of all, the CDW ansatz used here is only one possible configuration of an anisotropic chiral condensate (and the resulting anisotropic dispersion relations of the nucleons). For instance, it would be interesting to include the ``shifted'' ansatz of Ref.\ \cite{Takeda:2018ldi}, with the idea to possibly extend the region of the CDW into a regime with sufficiently stiff equations of state. Similarly, one can include a spatially modulated condensate in the charged pion sector \cite{Dautry:1979bk}. Beyond extensions to the CDW itself, which corresponds to a one-dimensional modulation, there is of course also the possibility of a three-dimensional crystalline structure, which possibly replaces the CDW solution. It would also be interesting to study the nuclear CDW in other models to alleviate the model dependence of our results. A candidate model is the extended linear sigma model of Ref.\ \cite{Heinz:2013hza}, where the CDW was studied without neutron star conditions and without fermionic vacuum fluctuations. While this is a phenomenological model in the same spirit as the one used here, a complementary strong-coupling approach, possibly using holographic methods
\cite{Kovensky:2021ddl,Kovensky:2021kzl,Kovensky:2023mye}, is also highly desired, as suggested by the large -- and in the present approximation uncontrolled -- effect of the vacuum fluctuations on the CDW. A more physical question concerns the presence of a magnetic field, which can be expected to favor anisotropic phases and one might ask which field strengths are needed to find a nuclear CDW in the core of neutron stars, similar to the magnetic CDW in a quark matter core discussed in Ref.\ \cite{Carignano:2015kda}. Finally, one could include strangeness into the model as done for isotropic phases in Ref.\ \cite{Fraga:2022yls}. This brings the chirally restored phase of the model closer to real-world quark matter and it would be interesting to see how this affects the conclusion about the CDW in neutron stars.

\section*{Acknowledgments}
We thank Eduardo Fraga, Alexander Haber, and Savvas Pitsinigkos for valuable comments and discussions.

\appendix

\section{Incompressibility, asymmetry energy, and slope parameter}
\label{app:KSL}

Here we calculate the incompressibility $K$, the asymmetry energy $S$, and the slope parameter $L$, defined by 
\begin{subequations}
\bea
K&=& 9n_B\frac{\partial \mu_B}{\partial n_B} \, , \label{Kdef} \\[2ex]
\label{Sdef}
S &=& 
\frac{n_B}{2} \frac{\partial\mu_I}{\partial n_I} \, , \\[2ex]
L &=& 3n_B\frac{\partial S}{\partial n_B} \, , \label{Ldef}
\eea
\end{subequations}
where the derivatives with respect to $n_B$ ($n_I$) are taken at fixed $n_I$ ($n_B$). After taking the derivatives, $n_I$ is set to zero in all three quantities and $n_B=n_0$, i.e., for our purpose they are defined in isospin-symmetric nuclear matter at saturation density (but can obviously be generalized to all densities). 

We start from the relation between the nucleon number density and the Fermi momentum,
\be
n_N=\frac{k_{F,N}^3}{3\pi^2} \, , \qquad k_{F,N}=\sqrt{(\mu_{N}^*)^2-M^2} \, , 
\ee
where $N=n,p$, 
which allows us to express the effective nucleon chemical potential as 
\be
\mu_{N}^* = \sqrt{(3\pi^2)^{2/3} n_{N}^{2/3}+M^2} \, .
\ee
Then, with the relation between the effective chemical potentials and the meson condensates $\omega$ and $\rho$ (\ref{mustar}), the baryon and isospin chemical potentials (\ref{muhat}) and densities (\ref{nBI})  we can write $\mu_B$ and $\mu_I$ as functions of $n_B$ and $n_I$ in the following way,
\begin{subequations}
\bea
\mu_B &=& g_\omega \omega+ \frac{1}{2}\left[\sqrt{\left(\frac{3\pi^2}{2}\right)^{2/3}(n_B+n_I)^{2/3}
+M^2}+\sqrt{\left(\frac{3\pi^2}{2}\right)^{2/3}(n_B-n_I)^{2/3}
+M^2}\right]\, , \\[2ex]
\mu_I  &=&  g_\rho \rho + \frac{1}{2}\left[\sqrt{\left(\frac{3\pi^2}{2}\right)^{2/3}(n_B+n_I)^{2/3}
+M^2}-\sqrt{\left(\frac{3\pi^2}{2}\right)^{2/3}(n_B-n_I)^{2/3}
+M^2}\right] \, ,
\eea
\end{subequations}
where $\omega$, $\rho$, and $M$ are (complicated) functions of $n_B$ and $n_I$, determined through the stationarity equations.  Hence, all possible derivatives are 
\begin{subequations}
\bea
\frac{\partial \mu_B}{\partial n_{B/I}} &=& g_\omega \frac{\partial\omega}{\partial n_{B/I}} + \frac{(3\pi^2)^{2/3}}{12}\left(\frac{1}{\mu_n^*n_n^{1/3}}\pm\frac{1}{\mu_p^*n_p^{1/3}}\right)+\left(\frac{1}{\mu_n^*}+\frac{1}{\mu_p^*}\right)\frac{M}{2}\frac{\partial M}{\partial n_{B/I}} \, , \\[2ex]
\frac{\partial \mu_I}{\partial n_{B/I}} &=& g_\rho \frac{\partial\rho}{\partial n_{B/I}} + \frac{(3\pi^2)^{2/3}}{12}\left(\frac{1}{\mu_n^*n_n^{1/3}}\mp\frac{1}{\mu_p^*n_p^{1/3}}\right)+\left(\frac{1}{\mu_n^*}-\frac{1}{\mu_p^*}\right)\frac{M}{2}\frac{\partial M}{\partial n_{B/I}} \, .
\label{dmuIdnI}
\eea
\end{subequations}
To compute the derivatives of $\omega$ and $\rho$ we need 
the stationarity equations (\ref{eq:omega}) and (\ref{eq:rho}), 
\begin{subequations}
\bea
0=f_1(\omega,\rho,n_B,n_I) &\equiv&  m_\omega^2\omega+(d_\omega\omega^2+d_{\omega\rho}\rho^2)\omega-g_\omega n_B  \, , \\[2ex]
0=f_2(\omega,\rho,n_B,n_I) &\equiv&m_\rho^2\rho+(d_{\omega\rho}\omega^2+d_{\rho}\rho^2)\rho -g_\rho n_I \, .
\eea
\end{subequations}
We can compute all relevant derivatives with the implicit function theorem,
\be
\left(\frac{\partial\omega}{\partial n},\frac{\partial\rho}{\partial n}\right) = -\left(\frac{\partial f_1}{\partial n},\frac{\partial f_2}{\partial n}\right)\left(\begin{array}{cc}\displaystyle{\frac{\partial f_1}{\partial \omega}} &  \displaystyle{\frac{\partial f_1}{\partial \rho}} \\[2ex] \displaystyle{\frac{\partial f_2}{\partial \omega}} &  \displaystyle{\frac{\partial f_2}{\partial \rho}}\end{array}\right)^{-1} \, ,
\ee
where $n$ stands for $n_B$ or $n_I$. We evaluate this to find
\begin{subequations}
\bea
\frac{\partial\omega}{\partial n_B} &=& \frac{g_\omega}{D}(m_\rho^2+d_{\omega\rho}\omega^2+3d_\rho\rho^2) \, , \\[2ex]
\frac{\partial\rho}{\partial n_B} &=& -\frac{2g_\omega}{D}d_{\omega\rho}\omega\rho  \, , \\[2ex]
\frac{\partial\omega}{\partial n_I} &=& -\frac{2g_\rho}{D}d_{\omega\rho}\omega\rho  \, , \\[2ex]
\frac{\partial\rho}{\partial n_I} &=& \frac{g_\rho}{D}(m_\omega^2+d_{\omega\rho}\rho^2+3d_\omega\omega^2) \, , \label{drhodnI}
\eea
\end{subequations}
where
\be
D\equiv (m_\omega^2+3d_\omega \omega^2 +d_{\omega\rho}\rho^2)(m_\rho^2+3d_\rho \rho^2 +d_{\omega\rho}\omega^2)-4d_{\omega\rho}^2\omega^2\rho^2 \, .
\ee
The incompressibility (\ref{Kdef}) is then straightforwardly computed as
\be \label{K1}
K = 9n_B\left[\frac{g_\omega^2}{m_\omega^2+3d_\omega \omega^2}+\frac{1}{3}\left(\frac{3\pi^2}{2}\right)^{2/3}\frac{1}{\mu_B^*n_B^{1/3}}+\frac{M}{\mu_B^*}\frac{\partial M}{\partial n_B}\right] \, , 
\ee
where we have set $\mu_n^*=\mu_p^*\equiv \mu_B^*$ and $\rho=0$ since we are interested in the result for symmetric nuclear matter. 
The derivative of $M$ with respect to $n_B$ can be obtained from the stationarity equation (\ref{eq:sigma})  (see for instance Appendix A of Ref.\ \cite{Schmitt:2020tac}),
\be
\frac{\partial M}{\partial n_B} = -\frac{M}{\mu_B^*}\left[\frac{\partial^2 \tilde{U}}{\partial M^2} +
\frac{1}{\pi^2}\left(\frac{k_F^3+3k_FM^2}{\mu_B^*}-3M^2\ln\frac{k_F+\mu_B^*}{M}\right)\right]^{-1} \, .
\ee
For the asymmetry energy (\ref{Sdef}) we find that the derivative of $M$ drops out for symmetric nuclear matter, 
and we obtain 
\be \label{S1}
S = \frac{g_\rho^2k_F^3}{3\pi^2(m_\rho^2+d_{\omega\rho}\omega^2)}+\frac{k_F^2}{6\mu_B^*} \, .
\ee
For the slope parameter $L$ (\ref{Ldef}) we compute $\frac{\partial S}{\partial n_B}$ and eliminate $\frac{\partial M}{\partial n_B}$ in favor of $K$ with the help of Eq.\ (\ref{K1}). This gives
\be\label{L1}
L = \frac{3n_B g_\rho^2}{2(m_\rho^2+d_{\omega\rho}\omega^2)}\left[1-\frac{2g_\omega d_{\omega\rho}\omega n_B}{(m_\rho^2+d_{\omega\rho}\omega^2)(m_\omega^2+3d_\omega \omega^2)}\right]+\frac{k_F^2}{3\mu_B^*}\left(1-\frac{K}{6\mu_B^*}\right)+\frac{g_\omega^2 n_B k_F^2}{2(\mu_B^*)^2(m_\omega^2+3d_\omega \omega^2)} \, .
\ee 
If we assume equal quartic vector meson couplings $d_\omega=d_\rho=d_{\omega\rho}$, this reduces to the result of Ref.\ \cite{Pitsinigkos:2023xee} (where the last term is written in a different, but equivalent, way).

\bibliography{references}

\begin{thebibliography}{70}
\expandafter\ifx\csname natexlab\endcsname\relax\def\natexlab#1{#1}\fi
\expandafter\ifx\csname bibnamefont\endcsname\relax
  \def\bibnamefont#1{#1}\fi
\expandafter\ifx\csname bibfnamefont\endcsname\relax
  \def\bibfnamefont#1{#1}\fi
\expandafter\ifx\csname citenamefont\endcsname\relax
  \def\citenamefont#1{#1}\fi
\expandafter\ifx\csname url\endcsname\relax
  \def\url#1{\texttt{#1}}\fi
\expandafter\ifx\csname urlprefix\endcsname\relax\def\urlprefix{URL }\fi
\providecommand{\bibinfo}[2]{#2}
\providecommand{\eprint}[2][]{\url{#2}}

\bibitem[{\citenamefont{Dautry and Nyman}(1979)}]{Dautry:1979bk}
\bibinfo{author}{\bibfnamefont{F.}~\bibnamefont{Dautry}} \bibnamefont{and}
  \bibinfo{author}{\bibfnamefont{E.~M.} \bibnamefont{Nyman}},
  \bibinfo{journal}{Nucl. Phys. A} \textbf{\bibinfo{volume}{319}},
  \bibinfo{pages}{323} (\bibinfo{year}{1979}).

\bibitem[{\citenamefont{Takahashi and Tatsumi}(2001)}]{Takahashi:2001jq}
\bibinfo{author}{\bibfnamefont{K.}~\bibnamefont{Takahashi}} \bibnamefont{and}
  \bibinfo{author}{\bibfnamefont{T.}~\bibnamefont{Tatsumi}},
  \bibinfo{journal}{Phys. Rev. C} \textbf{\bibinfo{volume}{63}},
  \bibinfo{pages}{015205} (\bibinfo{year}{2001}).

\bibitem[{\citenamefont{Takahashi}(2002)}]{Takahashi:2002md}
\bibinfo{author}{\bibfnamefont{K.}~\bibnamefont{Takahashi}},
  \bibinfo{journal}{Phys. Rev. C} \textbf{\bibinfo{volume}{66}},
  \bibinfo{pages}{025202} (\bibinfo{year}{2002}).

\bibitem[{\citenamefont{Buballa and Carignano}(2015)}]{Buballa:2014tba}
\bibinfo{author}{\bibfnamefont{M.}~\bibnamefont{Buballa}} \bibnamefont{and}
  \bibinfo{author}{\bibfnamefont{S.}~\bibnamefont{Carignano}},
  \bibinfo{journal}{Prog. Part. Nucl. Phys.} \textbf{\bibinfo{volume}{81}},
  \bibinfo{pages}{39} (\bibinfo{year}{2015}), \eprint{1406.1367}.

\bibitem[{\citenamefont{Alford et~al.}(2001)\citenamefont{Alford, Bowers, and
  Rajagopal}}]{Alford:2000ze}
\bibinfo{author}{\bibfnamefont{M.~G.} \bibnamefont{Alford}},
  \bibinfo{author}{\bibfnamefont{J.~A.} \bibnamefont{Bowers}},
  \bibnamefont{and}
  \bibinfo{author}{\bibfnamefont{K.}~\bibnamefont{Rajagopal}},
  \bibinfo{journal}{Phys. Rev. D} \textbf{\bibinfo{volume}{63}},
  \bibinfo{pages}{074016} (\bibinfo{year}{2001}), \eprint{hep-ph/0008208}.

\bibitem[{\citenamefont{Sch\"afer}(2006)}]{Schafer:2005ym}
\bibinfo{author}{\bibfnamefont{T.}~\bibnamefont{Sch\"afer}},
  \bibinfo{journal}{Phys. Rev. Lett.} \textbf{\bibinfo{volume}{96}},
  \bibinfo{pages}{012305} (\bibinfo{year}{2006}), \eprint{hep-ph/0508190}.

\bibitem[{\citenamefont{Kryjevski}(2008)}]{Kryjevski:2005qq}
\bibinfo{author}{\bibfnamefont{A.}~\bibnamefont{Kryjevski}},
  \bibinfo{journal}{Phys. Rev. D} \textbf{\bibinfo{volume}{77}},
  \bibinfo{pages}{014018} (\bibinfo{year}{2008}), \eprint{hep-ph/0508180}.

\bibitem[{\citenamefont{Alford et~al.}(2008)\citenamefont{Alford, Schmitt,
  Rajagopal, and Sch{\"a}fer}}]{Alford:2007xm}
\bibinfo{author}{\bibfnamefont{M.~G.} \bibnamefont{Alford}},
  \bibinfo{author}{\bibfnamefont{A.}~\bibnamefont{Schmitt}},
  \bibinfo{author}{\bibfnamefont{K.}~\bibnamefont{Rajagopal}},
  \bibnamefont{and}
  \bibinfo{author}{\bibfnamefont{T.}~\bibnamefont{Sch{\"a}fer}},
  \bibinfo{journal}{Rev.Mod.Phys.} \textbf{\bibinfo{volume}{80}},
  \bibinfo{pages}{1455} (\bibinfo{year}{2008}), \eprint{0709.4635}.

\bibitem[{\citenamefont{Pitsinigkos and Schmitt}(2024)}]{Pitsinigkos:2023xee}
\bibinfo{author}{\bibfnamefont{S.}~\bibnamefont{Pitsinigkos}} \bibnamefont{and}
  \bibinfo{author}{\bibfnamefont{A.}~\bibnamefont{Schmitt}},
  \bibinfo{journal}{Phys. Rev. D} \textbf{\bibinfo{volume}{109}},
  \bibinfo{pages}{014024} (\bibinfo{year}{2024}), \eprint{2309.01603}.

\bibitem[{\citenamefont{Nickel}(2009)}]{Nickel:2009wj}
\bibinfo{author}{\bibfnamefont{D.}~\bibnamefont{Nickel}},
  \bibinfo{journal}{Phys.Rev.} \textbf{\bibinfo{volume}{D80}},
  \bibinfo{pages}{074025} (\bibinfo{year}{2009}), \eprint{0906.5295}.

\bibitem[{\citenamefont{Boguta}(1983)}]{Boguta:1982wr}
\bibinfo{author}{\bibfnamefont{J.}~\bibnamefont{Boguta}},
  \bibinfo{journal}{Phys. Lett.} \textbf{\bibinfo{volume}{120B}},
  \bibinfo{pages}{34} (\bibinfo{year}{1983}).

\bibitem[{\citenamefont{Boguta and Kunz}(1989)}]{Boguta:1986ha}
\bibinfo{author}{\bibfnamefont{J.}~\bibnamefont{Boguta}} \bibnamefont{and}
  \bibinfo{author}{\bibfnamefont{J.}~\bibnamefont{Kunz}},
  \bibinfo{journal}{Nucl. Phys.} \textbf{\bibinfo{volume}{A501}},
  \bibinfo{pages}{637} (\bibinfo{year}{1989}).

\bibitem[{\citenamefont{Floerchinger and
  Wetterich}(2012)}]{Floerchinger:2012xd}
\bibinfo{author}{\bibfnamefont{S.}~\bibnamefont{Floerchinger}}
  \bibnamefont{and}
  \bibinfo{author}{\bibfnamefont{C.}~\bibnamefont{Wetterich}},
  \bibinfo{journal}{Nucl.Phys.} \textbf{\bibinfo{volume}{A890-891}},
  \bibinfo{pages}{11} (\bibinfo{year}{2012}), \eprint{1202.1671}.

\bibitem[{\citenamefont{Drews et~al.}(2013)\citenamefont{Drews, Hell, Klein,
  and Weise}}]{Drews:2013hha}
\bibinfo{author}{\bibfnamefont{M.}~\bibnamefont{Drews}},
  \bibinfo{author}{\bibfnamefont{T.}~\bibnamefont{Hell}},
  \bibinfo{author}{\bibfnamefont{B.}~\bibnamefont{Klein}}, \bibnamefont{and}
  \bibinfo{author}{\bibfnamefont{W.}~\bibnamefont{Weise}},
  \bibinfo{journal}{Phys. Rev. D} \textbf{\bibinfo{volume}{88}},
  \bibinfo{pages}{096011} (\bibinfo{year}{2013}), \eprint{1308.5596}.

\bibitem[{\citenamefont{Drews and Weise}(2015)}]{Drews:2014spa}
\bibinfo{author}{\bibfnamefont{M.}~\bibnamefont{Drews}} \bibnamefont{and}
  \bibinfo{author}{\bibfnamefont{W.}~\bibnamefont{Weise}},
  \bibinfo{journal}{Phys. Rev.} \textbf{\bibinfo{volume}{C91}},
  \bibinfo{pages}{035802} (\bibinfo{year}{2015}), \eprint{1412.7655}.

\bibitem[{\citenamefont{Fraga et~al.}(2019)\citenamefont{Fraga, Hippert, and
  Schmitt}}]{Fraga:2018cvr}
\bibinfo{author}{\bibfnamefont{E.~S.} \bibnamefont{Fraga}},
  \bibinfo{author}{\bibfnamefont{M.}~\bibnamefont{Hippert}}, \bibnamefont{and}
  \bibinfo{author}{\bibfnamefont{A.}~\bibnamefont{Schmitt}},
  \bibinfo{journal}{Phys. Rev. D} \textbf{\bibinfo{volume}{99}},
  \bibinfo{pages}{014046} (\bibinfo{year}{2019}), \eprint{1810.13226}.

\bibitem[{\citenamefont{Schmitt}(2020)}]{Schmitt:2020tac}
\bibinfo{author}{\bibfnamefont{A.}~\bibnamefont{Schmitt}},
  \bibinfo{journal}{Phys. Rev. D} \textbf{\bibinfo{volume}{101}},
  \bibinfo{pages}{074007} (\bibinfo{year}{2020}), \eprint{2002.01451}.

\bibitem[{\citenamefont{Fraga et~al.}(2022)\citenamefont{Fraga, da~Mata,
  Pitsinigkos, and Schmitt}}]{Fraga:2022yls}
\bibinfo{author}{\bibfnamefont{E.~S.} \bibnamefont{Fraga}},
  \bibinfo{author}{\bibfnamefont{R.}~\bibnamefont{da~Mata}},
  \bibinfo{author}{\bibfnamefont{S.}~\bibnamefont{Pitsinigkos}},
  \bibnamefont{and} \bibinfo{author}{\bibfnamefont{A.}~\bibnamefont{Schmitt}},
  \bibinfo{journal}{Phys. Rev. D} \textbf{\bibinfo{volume}{106}},
  \bibinfo{pages}{074018} (\bibinfo{year}{2022}), \eprint{2206.09219}.

\bibitem[{\citenamefont{Nakano and Tatsumi}(2005)}]{Nakano:2004cd}
\bibinfo{author}{\bibfnamefont{E.}~\bibnamefont{Nakano}} \bibnamefont{and}
  \bibinfo{author}{\bibfnamefont{T.}~\bibnamefont{Tatsumi}},
  \bibinfo{journal}{Phys.Rev.} \textbf{\bibinfo{volume}{D71}},
  \bibinfo{pages}{114006} (\bibinfo{year}{2005}), \eprint{hep-ph/0411350}.

\bibitem[{\citenamefont{Frolov et~al.}(2010)\citenamefont{Frolov, Zhukovsky,
  and Klimenko}}]{Frolov:2010wn}
\bibinfo{author}{\bibfnamefont{I.~E.} \bibnamefont{Frolov}},
  \bibinfo{author}{\bibfnamefont{V.~C.} \bibnamefont{Zhukovsky}},
  \bibnamefont{and} \bibinfo{author}{\bibfnamefont{K.~G.}
  \bibnamefont{Klimenko}}, \bibinfo{journal}{Phys.Rev.}
  \textbf{\bibinfo{volume}{D82}}, \bibinfo{pages}{076002}
  (\bibinfo{year}{2010}), \eprint{1007.2984}.

\bibitem[{\citenamefont{Carignano and
  Buballa}(2012{\natexlab{a}})}]{Carignano:2011gr}
\bibinfo{author}{\bibfnamefont{S.}~\bibnamefont{Carignano}} \bibnamefont{and}
  \bibinfo{author}{\bibfnamefont{M.}~\bibnamefont{Buballa}},
  \bibinfo{journal}{Acta Phys. Polon. Supp.} \textbf{\bibinfo{volume}{5}},
  \bibinfo{pages}{641} (\bibinfo{year}{2012}{\natexlab{a}}),
  \eprint{1111.4400}.

\bibitem[{\citenamefont{Carignano et~al.}(2014)\citenamefont{Carignano,
  Buballa, and Schaefer}}]{Carignano:2014jla}
\bibinfo{author}{\bibfnamefont{S.}~\bibnamefont{Carignano}},
  \bibinfo{author}{\bibfnamefont{M.}~\bibnamefont{Buballa}}, \bibnamefont{and}
  \bibinfo{author}{\bibfnamefont{B.-J.} \bibnamefont{Schaefer}},
  \bibinfo{journal}{Phys. Rev. D} \textbf{\bibinfo{volume}{90}},
  \bibinfo{pages}{014033} (\bibinfo{year}{2014}), \eprint{1404.0057}.

\bibitem[{\citenamefont{Buballa and Carignano}(2016)}]{Buballa:2015awa}
\bibinfo{author}{\bibfnamefont{M.}~\bibnamefont{Buballa}} \bibnamefont{and}
  \bibinfo{author}{\bibfnamefont{S.}~\bibnamefont{Carignano}},
  \bibinfo{journal}{Eur. Phys. J. A} \textbf{\bibinfo{volume}{52}},
  \bibinfo{pages}{57} (\bibinfo{year}{2016}), \eprint{1508.04361}.

\bibitem[{\citenamefont{Adhikari et~al.}(2017)\citenamefont{Adhikari, Andersen,
  and Kneschke}}]{Adhikari:2017ydi}
\bibinfo{author}{\bibfnamefont{P.}~\bibnamefont{Adhikari}},
  \bibinfo{author}{\bibfnamefont{J.~O.} \bibnamefont{Andersen}},
  \bibnamefont{and} \bibinfo{author}{\bibfnamefont{P.}~\bibnamefont{Kneschke}},
  \bibinfo{journal}{Phys. Rev. D} \textbf{\bibinfo{volume}{96}},
  \bibinfo{pages}{016013} (\bibinfo{year}{2017}), \bibinfo{note}{[Erratum:
  Phys.Rev.D 98, 099902 (2018)]}, \eprint{1702.01324}.

\bibitem[{\citenamefont{Andersen and Kneschke}(2018)}]{Andersen:2018osr}
\bibinfo{author}{\bibfnamefont{J.~O.} \bibnamefont{Andersen}} \bibnamefont{and}
  \bibinfo{author}{\bibfnamefont{P.}~\bibnamefont{Kneschke}},
  \bibinfo{journal}{Phys. Rev. D} \textbf{\bibinfo{volume}{97}},
  \bibinfo{pages}{076005} (\bibinfo{year}{2018}), \eprint{1802.01832}.

\bibitem[{\citenamefont{Ferrer and de~la Incera}(2020)}]{Ferrer:2019zfp}
\bibinfo{author}{\bibfnamefont{E.~J.} \bibnamefont{Ferrer}} \bibnamefont{and}
  \bibinfo{author}{\bibfnamefont{V.}~\bibnamefont{de~la Incera}},
  \bibinfo{journal}{Phys. Rev. D} \textbf{\bibinfo{volume}{102}},
  \bibinfo{pages}{014010} (\bibinfo{year}{2020}), \eprint{1902.06810}.

\bibitem[{\citenamefont{Carignano and Buballa}(2020)}]{Carignano:2019ivp}
\bibinfo{author}{\bibfnamefont{S.}~\bibnamefont{Carignano}} \bibnamefont{and}
  \bibinfo{author}{\bibfnamefont{M.}~\bibnamefont{Buballa}},
  \bibinfo{journal}{Phys. Rev. D} \textbf{\bibinfo{volume}{101}},
  \bibinfo{pages}{014026} (\bibinfo{year}{2020}), \eprint{1910.03604}.

\bibitem[{\citenamefont{Lakaschus et~al.}(2021)\citenamefont{Lakaschus,
  Buballa, and Rischke}}]{Lakaschus:2020caq}
\bibinfo{author}{\bibfnamefont{P.}~\bibnamefont{Lakaschus}},
  \bibinfo{author}{\bibfnamefont{M.}~\bibnamefont{Buballa}}, \bibnamefont{and}
  \bibinfo{author}{\bibfnamefont{D.~H.} \bibnamefont{Rischke}},
  \bibinfo{journal}{Phys. Rev. D} \textbf{\bibinfo{volume}{103}},
  \bibinfo{pages}{034030} (\bibinfo{year}{2021}), \eprint{2012.07520}.

\bibitem[{\citenamefont{Buballa et~al.}(2020)\citenamefont{Buballa, Carignano,
  and Kurth}}]{Buballa:2020xaa}
\bibinfo{author}{\bibfnamefont{M.}~\bibnamefont{Buballa}},
  \bibinfo{author}{\bibfnamefont{S.}~\bibnamefont{Carignano}},
  \bibnamefont{and} \bibinfo{author}{\bibfnamefont{L.}~\bibnamefont{Kurth}},
  \bibinfo{journal}{Eur. Phys. J. ST} \textbf{\bibinfo{volume}{229}},
  \bibinfo{pages}{3371} (\bibinfo{year}{2020}), \eprint{2006.02133}.

\bibitem[{\citenamefont{Heinz et~al.}(2015)\citenamefont{Heinz, Giacosa, and
  Rischke}}]{Heinz:2013hza}
\bibinfo{author}{\bibfnamefont{A.}~\bibnamefont{Heinz}},
  \bibinfo{author}{\bibfnamefont{F.}~\bibnamefont{Giacosa}}, \bibnamefont{and}
  \bibinfo{author}{\bibfnamefont{D.~H.} \bibnamefont{Rischke}},
  \bibinfo{journal}{Nucl. Phys. A} \textbf{\bibinfo{volume}{933}},
  \bibinfo{pages}{34} (\bibinfo{year}{2015}), \eprint{1312.3244}.

\bibitem[{\citenamefont{Takeda et~al.}(2018)\citenamefont{Takeda, Abuki, and
  Harada}}]{Takeda:2018ldi}
\bibinfo{author}{\bibfnamefont{Y.}~\bibnamefont{Takeda}},
  \bibinfo{author}{\bibfnamefont{H.}~\bibnamefont{Abuki}}, \bibnamefont{and}
  \bibinfo{author}{\bibfnamefont{M.}~\bibnamefont{Harada}},
  \bibinfo{journal}{Phys. Rev. D} \textbf{\bibinfo{volume}{97}},
  \bibinfo{pages}{094032} (\bibinfo{year}{2018}), \eprint{1803.06779}.

\bibitem[{\citenamefont{Abuki et~al.}(2012)\citenamefont{Abuki, Ishibashi, and
  Suzuki}}]{Abuki:2011pf}
\bibinfo{author}{\bibfnamefont{H.}~\bibnamefont{Abuki}},
  \bibinfo{author}{\bibfnamefont{D.}~\bibnamefont{Ishibashi}},
  \bibnamefont{and} \bibinfo{author}{\bibfnamefont{K.}~\bibnamefont{Suzuki}},
  \bibinfo{journal}{Phys. Rev. D} \textbf{\bibinfo{volume}{85}},
  \bibinfo{pages}{074002} (\bibinfo{year}{2012}), \eprint{1109.1615}.

\bibitem[{\citenamefont{Carignano and
  Buballa}(2012{\natexlab{b}})}]{Carignano:2012sx}
\bibinfo{author}{\bibfnamefont{S.}~\bibnamefont{Carignano}} \bibnamefont{and}
  \bibinfo{author}{\bibfnamefont{M.}~\bibnamefont{Buballa}},
  \bibinfo{journal}{Phys. Rev. D} \textbf{\bibinfo{volume}{86}},
  \bibinfo{pages}{074018} (\bibinfo{year}{2012}{\natexlab{b}}),
  \eprint{1203.5343}.

\bibitem[{\citenamefont{Pisarski et~al.}(2020)\citenamefont{Pisarski, Tsvelik,
  and Valgushev}}]{Pisarski:2020dnx}
\bibinfo{author}{\bibfnamefont{R.~D.} \bibnamefont{Pisarski}},
  \bibinfo{author}{\bibfnamefont{A.~M.} \bibnamefont{Tsvelik}},
  \bibnamefont{and}
  \bibinfo{author}{\bibfnamefont{S.}~\bibnamefont{Valgushev}},
  \bibinfo{journal}{Phys. Rev. D} \textbf{\bibinfo{volume}{102}},
  \bibinfo{pages}{016015} (\bibinfo{year}{2020}), \eprint{2005.10259}.

\bibitem[{\citenamefont{Winstel and Valgushev}(2024)}]{Winstel:2024qle}
\bibinfo{author}{\bibfnamefont{M.}~\bibnamefont{Winstel}} \bibnamefont{and}
  \bibinfo{author}{\bibfnamefont{S.}~\bibnamefont{Valgushev}}, in
  \emph{\bibinfo{booktitle}{{Excited QCD 2024 Workshop}}}
  (\bibinfo{year}{2024}), \eprint{2403.18640}.

\bibitem[{\citenamefont{Glendenning}(1992)}]{Glendenning:1992vb}
\bibinfo{author}{\bibfnamefont{N.~K.} \bibnamefont{Glendenning}},
  \bibinfo{journal}{Phys. Rev.} \textbf{\bibinfo{volume}{D46}},
  \bibinfo{pages}{1274} (\bibinfo{year}{1992}).

\bibitem[{\citenamefont{Heiselberg et~al.}(1993)\citenamefont{Heiselberg,
  Pethick, and Staubo}}]{PhysRevLett.70.1355}
\bibinfo{author}{\bibfnamefont{H.}~\bibnamefont{Heiselberg}},
  \bibinfo{author}{\bibfnamefont{C.~J.} \bibnamefont{Pethick}},
  \bibnamefont{and} \bibinfo{author}{\bibfnamefont{E.~F.}
  \bibnamefont{Staubo}}, \bibinfo{journal}{Phys. Rev. Lett.}
  \textbf{\bibinfo{volume}{70}}, \bibinfo{pages}{1355} (\bibinfo{year}{1993}).

\bibitem[{\citenamefont{Carignano et~al.}(2015)\citenamefont{Carignano, Ferrer,
  de~la Incera, and Paulucci}}]{Carignano:2015kda}
\bibinfo{author}{\bibfnamefont{S.}~\bibnamefont{Carignano}},
  \bibinfo{author}{\bibfnamefont{E.~J.} \bibnamefont{Ferrer}},
  \bibinfo{author}{\bibfnamefont{V.}~\bibnamefont{de~la Incera}},
  \bibnamefont{and} \bibinfo{author}{\bibfnamefont{L.}~\bibnamefont{Paulucci}},
  \bibinfo{journal}{Phys. Rev. D} \textbf{\bibinfo{volume}{92}},
  \bibinfo{pages}{105018} (\bibinfo{year}{2015}), \eprint{1505.05094}.

\bibitem[{\citenamefont{Ferrer and de~la Incera}(2021)}]{Ferrer:2021mpq}
\bibinfo{author}{\bibfnamefont{E.~J.} \bibnamefont{Ferrer}} \bibnamefont{and}
  \bibinfo{author}{\bibfnamefont{V.}~\bibnamefont{de~la Incera}},
  \bibinfo{journal}{Universe} \textbf{\bibinfo{volume}{7}},
  \bibinfo{pages}{458} (\bibinfo{year}{2021}), \eprint{2201.04032}.

\bibitem[{\citenamefont{Tatsumi and Muto}(2014)}]{Tatsumi:2014cea}
\bibinfo{author}{\bibfnamefont{T.}~\bibnamefont{Tatsumi}} \bibnamefont{and}
  \bibinfo{author}{\bibfnamefont{T.}~\bibnamefont{Muto}},
  \bibinfo{journal}{Phys. Rev. D} \textbf{\bibinfo{volume}{89}},
  \bibinfo{pages}{103005} (\bibinfo{year}{2014}), \eprint{1403.1927}.

\bibitem[{\citenamefont{Tews et~al.}(2018)\citenamefont{Tews, Carlson,
  Gandolfi, and Reddy}}]{Tews:2018kmu}
\bibinfo{author}{\bibfnamefont{I.}~\bibnamefont{Tews}},
  \bibinfo{author}{\bibfnamefont{J.}~\bibnamefont{Carlson}},
  \bibinfo{author}{\bibfnamefont{S.}~\bibnamefont{Gandolfi}}, \bibnamefont{and}
  \bibinfo{author}{\bibfnamefont{S.}~\bibnamefont{Reddy}},
  \bibinfo{journal}{Astrophys. J.} \textbf{\bibinfo{volume}{860}},
  \bibinfo{pages}{149} (\bibinfo{year}{2018}), \eprint{1801.01923}.

\bibitem[{\citenamefont{Alford et~al.}(2022)\citenamefont{Alford, Brodie,
  Haber, and Tews}}]{Alford:2022bpp}
\bibinfo{author}{\bibfnamefont{M.~G.} \bibnamefont{Alford}},
  \bibinfo{author}{\bibfnamefont{L.}~\bibnamefont{Brodie}},
  \bibinfo{author}{\bibfnamefont{A.}~\bibnamefont{Haber}}, \bibnamefont{and}
  \bibinfo{author}{\bibfnamefont{I.}~\bibnamefont{Tews}},
  \bibinfo{journal}{Phys. Rev. C} \textbf{\bibinfo{volume}{106}},
  \bibinfo{pages}{055804} (\bibinfo{year}{2022}), \eprint{2205.10283}.

\bibitem[{\citenamefont{Glendenning}(2000)}]{glendenning}
\bibinfo{author}{\bibfnamefont{N.~K.} \bibnamefont{Glendenning}},
  \emph{\bibinfo{title}{{Compact Stars}}} (\bibinfo{publisher}{Springer},
  \bibinfo{year}{2000}), ISBN \bibinfo{isbn}{978-0-387-98977-8}.

\bibitem[{\citenamefont{Steiner et~al.}(2005)\citenamefont{Steiner, Prakash,
  Lattimer, and Ellis}}]{Steiner:2004fi}
\bibinfo{author}{\bibfnamefont{A.~W.} \bibnamefont{Steiner}},
  \bibinfo{author}{\bibfnamefont{M.}~\bibnamefont{Prakash}},
  \bibinfo{author}{\bibfnamefont{J.~M.} \bibnamefont{Lattimer}},
  \bibnamefont{and} \bibinfo{author}{\bibfnamefont{P.~J.} \bibnamefont{Ellis}},
  \bibinfo{journal}{Phys. Rept.} \textbf{\bibinfo{volume}{411}},
  \bibinfo{pages}{325} (\bibinfo{year}{2005}), \eprint{nucl-th/0410066}.

\bibitem[{\citenamefont{Kapusta and Gale}(2023)}]{kapustabook}
\bibinfo{author}{\bibfnamefont{J.}~\bibnamefont{Kapusta}} \bibnamefont{and}
  \bibinfo{author}{\bibfnamefont{C.}~\bibnamefont{Gale}},
  \emph{\bibinfo{title}{Finite-temperature field theory: principles and
  applications}} (\bibinfo{publisher}{Cambridge University Press},
  \bibinfo{address}{Cambridge}, \bibinfo{year}{2023}), \bibinfo{edition}{2nd}
  ed.

\bibitem[{\citenamefont{Kurkela et~al.}(2010)\citenamefont{Kurkela, Romatschke,
  and Vuorinen}}]{Kurkela:2009gj}
\bibinfo{author}{\bibfnamefont{A.}~\bibnamefont{Kurkela}},
  \bibinfo{author}{\bibfnamefont{P.}~\bibnamefont{Romatschke}},
  \bibnamefont{and} \bibinfo{author}{\bibfnamefont{A.}~\bibnamefont{Vuorinen}},
  \bibinfo{journal}{Phys.Rev.} \textbf{\bibinfo{volume}{D81}},
  \bibinfo{pages}{105021} (\bibinfo{year}{2010}), \eprint{0912.1856}.

\bibitem[{\citenamefont{Fraga et~al.}(2023)\citenamefont{Fraga, Palhares, and
  Restrepo}}]{Fraga:2023cef}
\bibinfo{author}{\bibfnamefont{E.~S.} \bibnamefont{Fraga}},
  \bibinfo{author}{\bibfnamefont{L.~F.} \bibnamefont{Palhares}},
  \bibnamefont{and} \bibinfo{author}{\bibfnamefont{T.~E.}
  \bibnamefont{Restrepo}}, \bibinfo{journal}{Phys. Rev. D}
  \textbf{\bibinfo{volume}{108}}, \bibinfo{pages}{034026}
  (\bibinfo{year}{2023}), \eprint{2303.12140}.

\bibitem[{\citenamefont{Schmitt}(2010)}]{Schmitt:2010pn}
\bibinfo{author}{\bibfnamefont{A.}~\bibnamefont{Schmitt}},
  \bibinfo{journal}{Lect. Notes Phys.} \textbf{\bibinfo{volume}{811}},
  \bibinfo{pages}{1} (\bibinfo{year}{2010}), \eprint{1001.3294}.

\bibitem[{\citenamefont{Blaizot et~al.}(1995)\citenamefont{Blaizot, Berger,
  Decharge, and Girod}}]{Blaizot:1995zz}
\bibinfo{author}{\bibfnamefont{J.~P.} \bibnamefont{Blaizot}},
  \bibinfo{author}{\bibfnamefont{J.~F.} \bibnamefont{Berger}},
  \bibinfo{author}{\bibfnamefont{J.}~\bibnamefont{Decharge}}, \bibnamefont{and}
  \bibinfo{author}{\bibfnamefont{M.}~\bibnamefont{Girod}},
  \bibinfo{journal}{Nucl.Phys.} \textbf{\bibinfo{volume}{A591}},
  \bibinfo{pages}{435} (\bibinfo{year}{1995}).

\bibitem[{\citenamefont{Vretenar et~al.}(2003)\citenamefont{Vretenar, Niksic,
  and Ring}}]{Vretenar:2003qm}
\bibinfo{author}{\bibfnamefont{D.}~\bibnamefont{Vretenar}},
  \bibinfo{author}{\bibfnamefont{T.}~\bibnamefont{Niksic}}, \bibnamefont{and}
  \bibinfo{author}{\bibfnamefont{P.}~\bibnamefont{Ring}},
  \bibinfo{journal}{Phys. Rev. C} \textbf{\bibinfo{volume}{68}},
  \bibinfo{pages}{024310} (\bibinfo{year}{2003}), \eprint{nucl-th/0302070}.

\bibitem[{\citenamefont{Youngblood et~al.}(2004)\citenamefont{Youngblood, Lui,
  Clark, John, Tokimoto et~al.}}]{Youngblood:2004fe}
\bibinfo{author}{\bibfnamefont{D.~H.} \bibnamefont{Youngblood}},
  \bibinfo{author}{\bibfnamefont{Y.~W.} \bibnamefont{Lui}},
  \bibinfo{author}{\bibfnamefont{H.~L.} \bibnamefont{Clark}},
  \bibinfo{author}{\bibfnamefont{B.}~\bibnamefont{John}},
  \bibinfo{author}{\bibfnamefont{Y.}~\bibnamefont{Tokimoto}},
  \bibnamefont{et~al.}, \bibinfo{journal}{Phys.Rev.}
  \textbf{\bibinfo{volume}{C69}}, \bibinfo{pages}{034315}
  (\bibinfo{year}{2004}).

\bibitem[{\citenamefont{Shlomo et~al.}(2006)\citenamefont{Shlomo, Kolomietz,
  and Col\`o}}]{Shlomo:2006ole}
\bibinfo{author}{\bibfnamefont{S.}~\bibnamefont{Shlomo}},
  \bibinfo{author}{\bibfnamefont{V.~M.} \bibnamefont{Kolomietz}},
  \bibnamefont{and} \bibinfo{author}{\bibfnamefont{G.}~\bibnamefont{Col\`o}},
  \bibinfo{journal}{Eur. Phys. J. A} \textbf{\bibinfo{volume}{30}},
  \bibinfo{pages}{23} (\bibinfo{year}{2006}).

\bibitem[{\citenamefont{Liu et~al.}(2024)\citenamefont{Liu, Jiang, Chen, Wu,
  He, and Li}}]{Liu:2024qds}
\bibinfo{author}{\bibfnamefont{X.}~\bibnamefont{Liu}},
  \bibinfo{author}{\bibfnamefont{J.-D.} \bibnamefont{Jiang}},
  \bibinfo{author}{\bibfnamefont{X.}~\bibnamefont{Chen}},
  \bibinfo{author}{\bibfnamefont{X.-J.} \bibnamefont{Wu}},
  \bibinfo{author}{\bibfnamefont{B.}~\bibnamefont{He}}, \bibnamefont{and}
  \bibinfo{author}{\bibfnamefont{X.-H.} \bibnamefont{Li}},
  \bibinfo{journal}{Phys. Rev. C} \textbf{\bibinfo{volume}{110}},
  \bibinfo{pages}{034329} (\bibinfo{year}{2024}).

\bibitem[{\citenamefont{Skokov et~al.}(2010)\citenamefont{Skokov, Friman,
  Nakano, Redlich, and Schaefer}}]{Skokov:2010sf}
\bibinfo{author}{\bibfnamefont{V.}~\bibnamefont{Skokov}},
  \bibinfo{author}{\bibfnamefont{B.}~\bibnamefont{Friman}},
  \bibinfo{author}{\bibfnamefont{E.}~\bibnamefont{Nakano}},
  \bibinfo{author}{\bibfnamefont{K.}~\bibnamefont{Redlich}}, \bibnamefont{and}
  \bibinfo{author}{\bibfnamefont{B.~J.} \bibnamefont{Schaefer}},
  \bibinfo{journal}{Phys. Rev. D} \textbf{\bibinfo{volume}{82}},
  \bibinfo{pages}{034029} (\bibinfo{year}{2010}), \eprint{1005.3166}.

\bibitem[{\citenamefont{Adhikari et~al.}(2021)}]{PREX:2021umo}
\bibinfo{author}{\bibfnamefont{D.}~\bibnamefont{Adhikari}} \bibnamefont{et~al.}
  (\bibinfo{collaboration}{PREX}), \bibinfo{journal}{Phys. Rev. Lett.}
  \textbf{\bibinfo{volume}{126}}, \bibinfo{pages}{172502}
  (\bibinfo{year}{2021}), \eprint{2102.10767}.

\bibitem[{\citenamefont{Hu et~al.}(2022)}]{Hu:2021trw}
\bibinfo{author}{\bibfnamefont{B.}~\bibnamefont{Hu}} \bibnamefont{et~al.},
  \bibinfo{journal}{Nature Phys.} \textbf{\bibinfo{volume}{18}},
  \bibinfo{pages}{1196} (\bibinfo{year}{2022}), \eprint{2112.01125}.

\bibitem[{\citenamefont{Sotani et~al.}(2022)\citenamefont{Sotani, Nishimura,
  and Naito}}]{Sotani:2022hhq}
\bibinfo{author}{\bibfnamefont{H.}~\bibnamefont{Sotani}},
  \bibinfo{author}{\bibfnamefont{N.}~\bibnamefont{Nishimura}},
  \bibnamefont{and} \bibinfo{author}{\bibfnamefont{T.}~\bibnamefont{Naito}},
  \bibinfo{journal}{PTEP} \textbf{\bibinfo{volume}{2022}},
  \bibinfo{pages}{041D01} (\bibinfo{year}{2022}), \eprint{2203.05410}.

\bibitem[{\citenamefont{Adhikari et~al.}(2022)}]{CREX:2022kgg}
\bibinfo{author}{\bibfnamefont{D.}~\bibnamefont{Adhikari}} \bibnamefont{et~al.}
  (\bibinfo{collaboration}{CREX}), \bibinfo{journal}{Phys. Rev. Lett.}
  \textbf{\bibinfo{volume}{129}}, \bibinfo{pages}{042501}
  (\bibinfo{year}{2022}), \eprint{2205.11593}.

\bibitem[{\citenamefont{Lattimer}(2023)}]{Lattimer:2023rpe}
\bibinfo{author}{\bibfnamefont{J.~M.} \bibnamefont{Lattimer}},
  \bibinfo{journal}{Particles} \textbf{\bibinfo{volume}{6}},
  \bibinfo{pages}{30} (\bibinfo{year}{2023}), \eprint{2301.03666}.

\bibitem[{\citenamefont{Giacalone et~al.}(2023)\citenamefont{Giacalone, Nijs,
  and van~der Schee}}]{Giacalone:2023cet}
\bibinfo{author}{\bibfnamefont{G.}~\bibnamefont{Giacalone}},
  \bibinfo{author}{\bibfnamefont{G.}~\bibnamefont{Nijs}}, \bibnamefont{and}
  \bibinfo{author}{\bibfnamefont{W.}~\bibnamefont{van~der Schee}},
  \bibinfo{journal}{Phys. Rev. Lett.} \textbf{\bibinfo{volume}{131}},
  \bibinfo{pages}{202302} (\bibinfo{year}{2023}), \eprint{2305.00015}.

\bibitem[{\citenamefont{Tolman}(1934)}]{tolman}
\bibinfo{author}{\bibfnamefont{R.~C.} \bibnamefont{Tolman}},
  \emph{\bibinfo{title}{{Relativity, Thermodynamics and Cosmology}}}
  (\bibinfo{publisher}{Oxford University Press}, \bibinfo{year}{1934}).

\bibitem[{\citenamefont{Tolman}(1939)}]{PhysRev.55.364}
\bibinfo{author}{\bibfnamefont{R.~C.} \bibnamefont{Tolman}},
  \bibinfo{journal}{Phys. Rev.} \textbf{\bibinfo{volume}{55}},
  \bibinfo{pages}{364} (\bibinfo{year}{1939}).

\bibitem[{\citenamefont{Oppenheimer and Volkoff}(1939)}]{PhysRev.55.374}
\bibinfo{author}{\bibfnamefont{J.~R.} \bibnamefont{Oppenheimer}}
  \bibnamefont{and} \bibinfo{author}{\bibfnamefont{G.~M.}
  \bibnamefont{Volkoff}}, \bibinfo{journal}{Phys. Rev.}
  \textbf{\bibinfo{volume}{55}}, \bibinfo{pages}{374} (\bibinfo{year}{1939}).

\bibitem[{\citenamefont{Kovensky et~al.}(2022)\citenamefont{Kovensky, Poole,
  and Schmitt}}]{Kovensky:2021kzl}
\bibinfo{author}{\bibfnamefont{N.}~\bibnamefont{Kovensky}},
  \bibinfo{author}{\bibfnamefont{A.}~\bibnamefont{Poole}}, \bibnamefont{and}
  \bibinfo{author}{\bibfnamefont{A.}~\bibnamefont{Schmitt}},
  \bibinfo{journal}{Phys. Rev. D} \textbf{\bibinfo{volume}{105}},
  \bibinfo{pages}{034022} (\bibinfo{year}{2022}), \eprint{2111.03374}.

\bibitem[{\citenamefont{{Bardeen} et~al.}(1966)\citenamefont{{Bardeen},
  {Thorne}, and {Meltzer}}}]{1966ApJ...145..505B}
\bibinfo{author}{\bibfnamefont{J.~M.} \bibnamefont{{Bardeen}}},
  \bibinfo{author}{\bibfnamefont{K.~S.} \bibnamefont{{Thorne}}},
  \bibnamefont{and} \bibinfo{author}{\bibfnamefont{D.~W.}
  \bibnamefont{{Meltzer}}}, \bibinfo{journal}{\apj}
  \textbf{\bibinfo{volume}{145}}, \bibinfo{pages}{505} (\bibinfo{year}{1966}).

\bibitem[{\citenamefont{Fonseca et~al.}(2021)}]{Fonseca:2021wxt}
\bibinfo{author}{\bibfnamefont{E.}~\bibnamefont{Fonseca}} \bibnamefont{et~al.},
  \bibinfo{journal}{Astrophys. J. Lett.} \textbf{\bibinfo{volume}{915}},
  \bibinfo{pages}{L12} (\bibinfo{year}{2021}), \eprint{2104.00880}.

\bibitem[{\citenamefont{Romani et~al.}(2022)\citenamefont{Romani, Kandel,
  Filippenko, Brink, and Zheng}}]{Romani:2022jhd}
\bibinfo{author}{\bibfnamefont{R.~W.} \bibnamefont{Romani}},
  \bibinfo{author}{\bibfnamefont{D.}~\bibnamefont{Kandel}},
  \bibinfo{author}{\bibfnamefont{A.~V.} \bibnamefont{Filippenko}},
  \bibinfo{author}{\bibfnamefont{T.~G.} \bibnamefont{Brink}}, \bibnamefont{and}
  \bibinfo{author}{\bibfnamefont{W.}~\bibnamefont{Zheng}},
  \bibinfo{journal}{Astrophys. J. Lett.} \textbf{\bibinfo{volume}{934}},
  \bibinfo{pages}{L17} (\bibinfo{year}{2022}), \eprint{2207.05124}.

\bibitem[{\citenamefont{Christian et~al.}(2018)\citenamefont{Christian, Zacchi,
  and Schaffner-Bielich}}]{Christian:2017jni}
\bibinfo{author}{\bibfnamefont{J.-E.} \bibnamefont{Christian}},
  \bibinfo{author}{\bibfnamefont{A.}~\bibnamefont{Zacchi}}, \bibnamefont{and}
  \bibinfo{author}{\bibfnamefont{J.}~\bibnamefont{Schaffner-Bielich}},
  \bibinfo{journal}{Eur. Phys. J. A} \textbf{\bibinfo{volume}{54}},
  \bibinfo{pages}{28} (\bibinfo{year}{2018}), \eprint{1707.07524}.

\bibitem[{\citenamefont{Kovensky and Schmitt}(2021)}]{Kovensky:2021ddl}
\bibinfo{author}{\bibfnamefont{N.}~\bibnamefont{Kovensky}} \bibnamefont{and}
  \bibinfo{author}{\bibfnamefont{A.}~\bibnamefont{Schmitt}},
  \bibinfo{journal}{SciPost Phys.} \textbf{\bibinfo{volume}{11}},
  \bibinfo{pages}{029} (\bibinfo{year}{2021}), \eprint{2105.03218}.

\bibitem[{\citenamefont{Kovensky et~al.}(2023)\citenamefont{Kovensky, Poole,
  and Schmitt}}]{Kovensky:2023mye}
\bibinfo{author}{\bibfnamefont{N.}~\bibnamefont{Kovensky}},
  \bibinfo{author}{\bibfnamefont{A.}~\bibnamefont{Poole}}, \bibnamefont{and}
  \bibinfo{author}{\bibfnamefont{A.}~\bibnamefont{Schmitt}},
  \bibinfo{journal}{SciPost Phys.} \textbf{\bibinfo{volume}{15}},
  \bibinfo{pages}{162} (\bibinfo{year}{2023}), \eprint{2302.10675}.

\end{thebibliography}

\end{document}